\documentclass[twocolumn,english]{revtex4-2}
\usepackage[T1]{fontenc}
\usepackage[latin9]{inputenc}
\usepackage{array}
\usepackage{multirow}
\usepackage{amsmath}
\usepackage{graphicx}

\makeatletter

\providecommand{\tabularnewline}{\\}

\makeatother

\usepackage{babel}
\begin{document}
\title{Inspecting molecular aggregate quadratic vibronic coupling effects
using squeezed coherent states}
\author{Mantas Jaku\v{c}ionis\textsuperscript{1}, Agnius Žukas\textsuperscript{1},
Darius Abramavi\v{c}ius\textsuperscript{1}}
\affiliation{\textsuperscript{1}Institute of Chemical Physics, Vilnius University,
Sauletekio Ave. 9-III, LT-10222 Vilnius, Lithuania}
\begin{abstract}
We present systematic comparison of three quantum mechanical approaches
describing excitation dynamics in molecular complexes using the Time-Dependent
variational principle (TDVP) with three increasing sophistication
trial wavefunctions (ansatze): Davydov $\text{D}_{2}$, squeezed $\text{D}_{2}$
($\text{sqD}_{2}$) and a numerically exact multiple $\text{D}_{2}$
($\text{mD}_{2}$) ansatz in order to characterize validity of the
$\text{sqD}_{2}$ ansatze. Numerical simulation of molecular aggregate
absorption and fluorescence spectra with intra- and intermolecular
vibrational modes, including quadratic electron-vibrational (vibronic)
coupling term, which is due to vibrational frequency shift upon pigment
excitation is presented. Simulated absorption and fluorescence spectra
of J type molecular dimer with high frequency intramolecular vibrational
modes obtained with $\text{D}_{2}$ and $\text{sqD}_{2}$ ansatze
matches spectra of $\text{mD}_{2}$ ansatz only in the single pigment
model without quadratic vibronic coupling. In general, the use of
$\text{mD}_{2}$ ansatz is required to model accurate dimer and larger
aggregate's spectra. For a J dimer aggregate coupled to a low frequency
intermolecular phonon bath, absorption and fluorescence spectra are
qualitatively similar using all three ansatze. The quadratic vibronic
coupling term in both absorption and fluorescence spectra manifests
itself as a lineshape peak amplitude redistribution, static frequency
shift and an additional shift, which is temperature dependent. Overall
the squeezed $\text{D}_{2}$ model does not result in considerable
improvement of simulation results compared to the simplest Davydov
$\text{D}_{2}$ approach.
\end{abstract}
\maketitle

\section{Introduction}

A fundamental aspect of the physics of optically excited molecules
and their complexes is the transport of excitation energy. Electronic
and vibronic couplings are two aspects that are crucial to this process
\citep{Valkunas2013a}. Complex quantum dynamics of electronic and
vibrational excitations are produced as a result of intermolecular
interactions right after the optical excitation. Their interplay is
essential for effective photosynthetic machinery in a natural setting
where the energy transfer, relaxation, and charge transfer play a
crucial role in initial stages of solar energy conversion \citep{Blankenship2008,Amerongen2010}.

The wavefunction-based TDVP method can be used to simulate molecular
aggregate excitation dynamics as well as their optical spectra with
respect to an ansatz (or parameterization form), which should be sufficiently
sophisticated to describe the aggregate's essential vibronic features.
One family of wavefunctions is called Davydov's ansatze \citep{Davydov1979,Scott1991,Zhao2021},
which utilize Gaussian wavepackets, also known as coherent states
(CS), to represent vibronic states of molecular aggregate. It has
been extensively used to compute spectra of molecules as well as to
examine excitation relaxation dynamics in single molecules and their
molecular aggregates \citep{Sun2010b,Chorosajev2016c,Jakucionis2018b,Jakucionis2020a,Sun2015a,Zhou2016a,Chorosajev2017b,Jakucionis2022}.

The trial wavefunction's selection greatly influences how accurate
the method is. It has been shown, that in some cases, for precise
modeling of molecular aggregates, the $D_{2}$ ansatz falls short
\citep{Zhou2016}, however, accuracy of vibrational mode representation
can be improved by expanding the available parameter space. The most
potent approach is to consider a superposition of multiple $\text{D}_{2}$
ansatze, known as the multi-Davydov $\text{D}_{2}$ ansatz. It considerably
increases accuracy, making TDVP with $\text{mD}_{2}$ a numerically
exact method. Spin-boson models \citep{Wang2016}, nonadiabatic dynamics
of molecules' dynamics \citep{Chen2019a,Jakucionis2020a}, linear
and nonlinear spectra of molecular aggregates \citep{Sun2015a,Zhou2016,MantasJakucionis2022}
have all been investigated using TDVP with $\text{mD}_{2}$.

Instead of considering superposition of ansatze, which is equivalent
to complete quantum treatment, one can expand available state space
of the $\text{D}_{2}$ ansatz incrementally. One approach is to replace
the CS with squeezed coherent states (sqCS), which has additional
degrees of freeedom (DOFs) which allow for wavepacket to contract
and expand along coordinate and momentum axes in it's phase space.
Presumably this should allow sqCS to better represent complicated
structure of realistic vibrational mode wavepackets, which become
non-Gaussian due to both electronic \citep{MantasJakucionis2022}
and quadratic vibronic \citep{Chorosajev2017b,Abramavicius2018e,Jakucionis2020a}
couplings.

In this work, we aim to compare accuracy of TDVP with three increasing
sophistication ansatze: the regular Davydov $\text{D}_{2}$, $\text{sqD}_{2}$
with sqCS and an exact $\text{mD}_{2}$ ansatz by analysing simulated
absorption and fluorescence spectra of a J-type dimer couped to high
frequency (intra-) and low freqency intermolecular vibrational modes.
In addition, we also consider the quadratic vibronic coupling term,
which induce wavepacket non-Gaussianity.

The rest of the paper is organized as follows: in Subsection II.A
we describe quadratic vibronic molecular aggregate model, considered
ansatze and shortly mention an approach to include finite temperature
into the model. In Subsection II.B we present theory of absorption
and fluorescence spectra using TDVP approach. In Section III we analyze
and compare J aggregate absorption and fluorescence spectra in three
vibrational mode regimes. Results are discussed and conclusions are
given in Section IV.

\section{Theory}

\global\long\def\cm{\text{cm}^{-1}}%

\subsection{Electron-vibrational molecular aggregate model theory\label{subsec:theoryA}}

The generic model system is a molecular aggregate made of $N$ chromophores
with resonant interaction between them. Each chromophore corresponds
to a single pigment molecule (site) which is a two-level electronic
quantum system with ground and excited states. Moreover, each pigment
is coupled to a set of vibrational degrees of freedom (DOF) corresponding
to either intra- or intermolecular vibrational modes. Vibrations are
explicitly modeled by quantum harmonic oscillators (QHO). The total
system Hamiltonian can then be written as \citep{Valkunas2013a,Bardeen2014,Amerongen2010,Schroter2015}

\begin{equation}
\hat{H}=\hat{H}_{\text{S}}+\hat{H}_{\text{V}}+\hat{H}_{\text{S-V}}+\hat{H}_{\text{S-V}^{2}},\label{eq:H}
\end{equation}
where $\hat{H}_{\text{S}}$ represents site Hamiltonian, $\hat{H}_{\text{V}}$
is a vibrational Hamiltonian, $\hat{H}_{\text{S-V}}$ is a first-order
interaction term between sites and vibrational modes, and $\hat{H}_{\text{S-V}^{2}}$
is the quadratic site-vibration coupling term. All of the above are
explicitly expressed as

\begin{align}
\hat{H}_{\text{S}}= & \sum_{n}\varepsilon_{n}\hat{a}_{n}^{\dagger}\hat{a}_{n}+\sum_{n,m}^{n\neq m}V_{nm}\hat{a}_{n}^{\dagger}\hat{a}_{m},\label{eq:H-s}\\
\hat{H}_{\text{V}}= & \sum_{k,q}\omega_{kq}^{g}\hat{b}_{kq}^{\dagger}\hat{b}_{kq},\label{eq:H-v}\\
\hat{H}_{\text{S-V}}= & -\sum_{n}\hat{a}_{n}^{\dagger}\hat{a}_{n}\sum_{q}\omega_{nq}^{e}f_{nq}\left(\hat{b}_{nq}^{\dagger}+\hat{b}_{nq}\right),\label{eq:H-s-v}\\
\hat{H}_{\text{\ensuremath{\text{S-V}^{2}}}}= & \frac{1}{4}\sum_{n}\hat{a}_{n}^{\dagger}\hat{a}_{n}\sum_{q}\left(w_{nq}^{e}-w_{nq}^{g}\right)\left(\hat{b}_{nq}^{\dagger}+\hat{b}_{nq}\right)^{2},\label{eq:H-nonlinear}
\end{align}
where $\varepsilon_{n}$ denotes the $n$th site electronic excitation
energy, whichincludes molecular reorganization energy, equal to $\Lambda_{n}=\sum_{q}\omega_{nq}^{e}f_{nq}^{2}$.
where summation index $q$ runs over vibrational modes. $V_{nm}$
is the resonant coupling between the $n$th and $m$th site, $\hat{a}_{n}^{\dagger}$
$\left(\hat{a}_{n}\right)$ are the creation (annihilation) operators
of chromophore electronic excitation, $\hat{b}_{nq}^{\dagger}$ $\left(\hat{b}_{nq}\right)$
are creation (annihilation) operators of vibrational excitations.
The linear vibronic coupling strength is given by dimensionless amplitude
$f_{nq}$. The quadratic vibronic coupling term, $\hat{H}_{\text{\ensuremath{\text{S-V}^{2}}}}$,
becomes relevant once the vibrational mode frequencies in electronic
ground state, $\omega_{nq}^{g}$, are different from the ones in excited
state, $\omega_{nq}^{e}$, otherwise this term does not contribute
\citep{Chorosajev2017b,Steffen2013,Tanimura2013,Zhang2020,Hu1993,Xu2018,Jakucionis2020a}.

To obtain linear absorption and fluorescence spectrum of the presented
vibronic model, we will be using the TDVP method, which will be applied
to three parameterized wavefunction ansatze with increasing sophistication.
All of them are based on the Davydov $\text{D}_{2}$ ansatz. First
of, the least sophisticated ansatz we will be testing, is the Davydov
$\text{D}_{2}$ ansatz. It considers a superposition of singly excited
aggregate configurations $|n\rangle=|1\rangle_{n}\prod_{m\neq n}|0\rangle$
\citep{Frenkel1931,Valkunas2013a}, with time-dependent amplitudes
$\alpha_{n}\left(t\right)$, while vibrational QHO states are expanded
in terms of CS. These are obtained by applying the translation operator
\begin{equation}
\hat{D}\left(\lambda_{kq}\left(t\right)\right)=\exp\left(\lambda_{kq}\left(t\right)\hat{b}_{kq}^{\dagger}-\text{h.c.}\right),
\end{equation}
with complex time-dependent displacement parameters, $\lambda_{kq}$,
to the QHO vacuum state denoted by $|0\rangle_{kq}$. Then the $\text{D}_{2}$
ansatz is defined as
\begin{equation}
|\Psi_{\text{D}_{2}}\left(t\right)\rangle=\sum_{n}\alpha_{n}\left(t\right)|n\rangle\prod_{k,q}|\lambda_{kq}\left(t\right)\rangle.
\end{equation}

In order to increase the complexity of ansatz to better represent
a complicated vibronic model states, in addition to the translation
operator, we can additionally apply the squeeze operator
\begin{equation}
\hat{S}\left(\zeta_{kq}\left(t\right)\right)=\exp\left(\frac{1}{2}\left(\zeta_{kq}^{*}\left(t\right)\hat{b}_{kq}^{2}-\text{h.c.}\right)\right),
\end{equation}
with complex-valued squeeze parameter $\zeta_{kq}(t)$, which squeezes
the Gaussian wavepacket and only then shifts the resulting squeezed
state along the coordinate and momentum axes. The resulting state
\begin{equation}
\hat{D}\left(\lambda_{kq}\left(t\right)\right)\hat{S}\left(\zeta_{kq}\left(t\right)\right)|0\rangle_{kq}=|\lambda_{kq}\left(t\right),\zeta_{kq}\left(t\right)\rangle,
\end{equation}
 is called a sqCS. For convenience, we express complex squeeze parameter
$\zeta_{kq}(t)$ in its polar form $\zeta_{kq}(t)=r_{kq}(t)e^{\text{i}\theta_{kq}(t)}$
where squeeze amplitude $r_{kq}(t)$ and squeeze angle $\theta_{kq}(t)$
are now real time-dependent parameters. Then the squeezed $\text{sqD}_{2}$
ansatz is defined as

\begin{equation}
|\Psi_{\text{sqD}_{2}}\left(t\right)\rangle=\sum_{n}\alpha_{n}\left(t\right)|n\rangle\prod_{k,q}|\lambda_{kq}\left(t\right),\zeta_{kq}\left(t\right)\rangle.
\end{equation}

Even more general approach to constructing the ansatz is to consider
a superposition of multiple copies of the $\text{D}_{2}$ ansatz.
It has been termed by the multiple Davydov $\text{D}_{2}$, $\text{mD}_{2}$
ansatz, and is defined as 
\begin{equation}
|\Psi_{\text{mD}_{2}}\left(t\right)\rangle=\sum_{i=1}^{M}\left(\sum_{n}\alpha_{i,n}\left(t\right)|n\rangle\prod_{k,q}|\lambda_{i,kq}\left(t\right)\rangle\right),
\end{equation}
where each $i-$th multiple corresponds to a superposition of electronic
state excitations accompanied by the vibrational state of an aggregate.
By increasing the number of multiples considered, $M$, ansatz state
space is expanded accordingly. Note, that $\text{mD}_{2}$ ansatz
with $M=1$ simplifies to the $\text{D}_{2}$ ansatz, while an arbitrary
wavefunction can be expressed when $M\to\infty$, making the approach
exact.

Time evolution of considered ansatze are obtained by solving their
respective equations of motion (EOM), which are given Appendix \ref{sec:EOM}.
A more in depth discussion of $\text{mD}_{2}$ ansatz EOM numerical
implementation can be found in Refs. \citep{MantasJakucionis2022,Werther2020}.

Inclusion of additional statistical physics concepts are required
in order to simulate finite temperature of the model. The thermal
ensemble will be constructed by considering independent wavefunction
trajectories $\gamma$, each with different initial conditions, and
thus energies. Notice that time propagation of wavefunction fully
conserves the total energy of each trajectory.

Considering excitation process, prior to molecular aggregate excitation
via an external field, the aggregate is in its electronic ground state
$|0\rangle$, while vibrational DOFs are thermally excited. Thus QHO
modes follow statistics of the canonical ensemble with respect to
aggregate ground electronic state. Characterization of the vibrational
manifold is straightfoward because all oscillators in electronic ground
state state are uncoupled. Diagonal density operator of a single QHO
can be written in the basis of CS with quasiprobability distribution
function \citep{Glauber1963,Chorosajev2016c,Wang2017b,Xie2017}
\begin{equation}
\mathcal{P}^{\left(\text{g/e}\right)}\left(\lambda\right)=\mathcal{Z}^{-1}\exp\left(-\left|\lambda\right|^{2}\left(\text{e}^{\frac{\omega_{g/e}}{k_{\text{B}}T}}-1\right)\right),\label{eq:Sudarshan-P-1}
\end{equation}
where $\mathcal{Z}$ is the partition function of QHO, $k_{\text{B}}$
is the Boltzmann constant and $T$ is the temperature. By sampling
$\mathcal{P}^{\left(\text{g}\right)}$ distribution, ground state
vibrational mode initial displacements $\lambda\left(0\right)$ are
obtained. Then, by taking average of observable $A$ over ensemble
of trajectories $\gamma$, one obtains thermally averaged observable.

In the case of $\text{D}_{2}$ ansatz, distributions $\mathcal{P}^{\left(\text{g}\right)}$
fully describe CS initial displacements without ambiguity. For the
$\text{sqD}_{2}$ ansatz, we again sample $\mathcal{P}^{\left(\text{g}\right)}$
to deduce displacements $\lambda_{kq}\left(0\right)$ and set the
squeeze parameters to $r_{kq}=1$, $\theta_{kq}=0$ (no squeezing).
This is still complete description of thermal equilibrium state due
to eq. \ref{eq:Sudarshan-P-1}. Lastly, in the case of $\text{mD}_{2}$
ansatz, we have $M$ equivalent ways to set $\lambda_{i,kq}\left(0\right)$
values. Therefore, we choose to initially populate the first multiple,
$i=1$, according to values sampled from $\mathcal{P}^{\left(\text{g}\right)}$,
and set the rest, $i\neq1$, terms to $\lambda_{i\neq1}\left(0\right)=0$
\citep{MantasJakucionis2022}.

\subsection{Absorption and fluorescence spectra theory using TDVP\label{subsec:Absorption-and-fluorescence theory}}

Two spectroscopic signals, the linear absorption and fluorescence
are the most widely employed spectroscopy tools used to infer information
on molecular systems. Assuming that the lifetime of excited state
is longer than the excited state thermal equilibration, it is well
known \citep{Mukamel1995a,Valkunas2013a} that the absorption/fluorescence
spectrum can be obtained by taking Fourier transform of the corresponding
time domain response function
\begin{equation}
A_{\text{abs/flor}}\left(\omega\right)=\text{Re}\int_{0}^{\infty}\text{d}t\text{e}^{i\omega t-\gamma_{\text{dep}}t}S_{\text{abs/flor}}^{\left(1\right)}\left(t\right).\label{eq:Abs}
\end{equation}
In the rotating wave and instantaneous aggregate-field interaction
approximations \citep{Mukamel1995a,MantasJakucionis2022}, the absorption-related
response function is given by linear response 
\begin{equation}
S_{\text{abs}}^{\left(1\right)}\left(t\right)=\frac{1}{\Gamma}\sum_{\gamma=1}^{\Gamma}\langle\Psi^{\left(\text{g}\right)}\left(0\right)|_{\gamma}\hat{\mu}_{-}e^{i\hat{H}t}\hat{\mu}_{+}e^{-i\hat{H}_{\text{G}}t}|\Psi^{\left(\text{g}\right)}\left(0\right)\rangle_{\gamma},\label{eq:S1}
\end{equation}
where the ground state Hamiltonian is equal to $\hat{H}_{\text{G}}=\hat{H}_{\text{V}}$.
Sum over $\gamma$ trajectories describe ensemble averaging over incoherent
ensemble of electronic ground states $|\Psi^{\left(\text{g}\right)}\left(0\right)\rangle_{\gamma}$
(for all ansatze) before excitation via the external field, where
each trajectory has different initial bath conditions, as described
previously in Section (\ref{subsec:theoryA}). $\Gamma$ is the total
number of trajectories of thermal ensemble.
\begin{align}
\hat{\mu}_{+} & =\sum_{n}\left(\boldsymbol{e}\cdot\boldsymbol{\mu}_{n}\right)\hat{a}_{n}^{\dagger},\label{eq:dipUP}\\
\hat{\mu}_{-} & =\sum_{n}\left(\boldsymbol{e}\cdot\boldsymbol{\mu}_{n}\right)\hat{a}_{n},\label{eq:dipDOWN}
\end{align}
are the aggregate excitation and deexcitation operators, $\boldsymbol{e}$
is the external field polarization vector, $\boldsymbol{\mu}_{n}$
is the $n$th molecule electronic transition dipole vector. In Eq.
(\ref{eq:Abs}) we include phenomenological dephasing rate, $\gamma_{\text{dep}}$,
to account for decay of coherence due to explicitly unaccounted dephasing
effects.

To describe fluorescence response function $S_{\text{flor}}^{\left(1\right)}\left(t\right)$,
a more general, third-order, time-resolved fluorescence (TRF) response
function \citep{Mukamel1995a,Balevicius2015c}
\begin{align}
S_{\text{trf}}^{\left(3\right)}\left(\tau,t\right) & =\frac{1}{\Gamma}\sum_{\gamma=1}^{\Gamma}\langle\Psi^{\left(\text{g}\right)}\left(0\right)|_{\gamma}\hat{\mu}_{+}e^{-i\hat{H}\left(\tau+t\right)}\hat{\mu}_{-}\nonumber \\
 & \times e^{-i\hat{H}_{\text{G}}t}\hat{\mu}_{-}e^{-i\hat{H}\tau}\hat{\mu}_{+}|\Psi^{\left(\text{g}\right)}\left(0\right)\rangle_{\gamma},\label{eq:trf_full}
\end{align}
must be used. Initially, first two aggregate-field interactions create
nonequilibrium density matrix configuration among electronic excited
states. Then the aggregate evolves for waiting time, $\tau$, after
which, deexcitation transition takes place by spontaneus emission
from the excited to the ground electronic state, defined by delay
time interval, $t$.
\begin{figure*}
\includegraphics[width=0.8\textwidth]{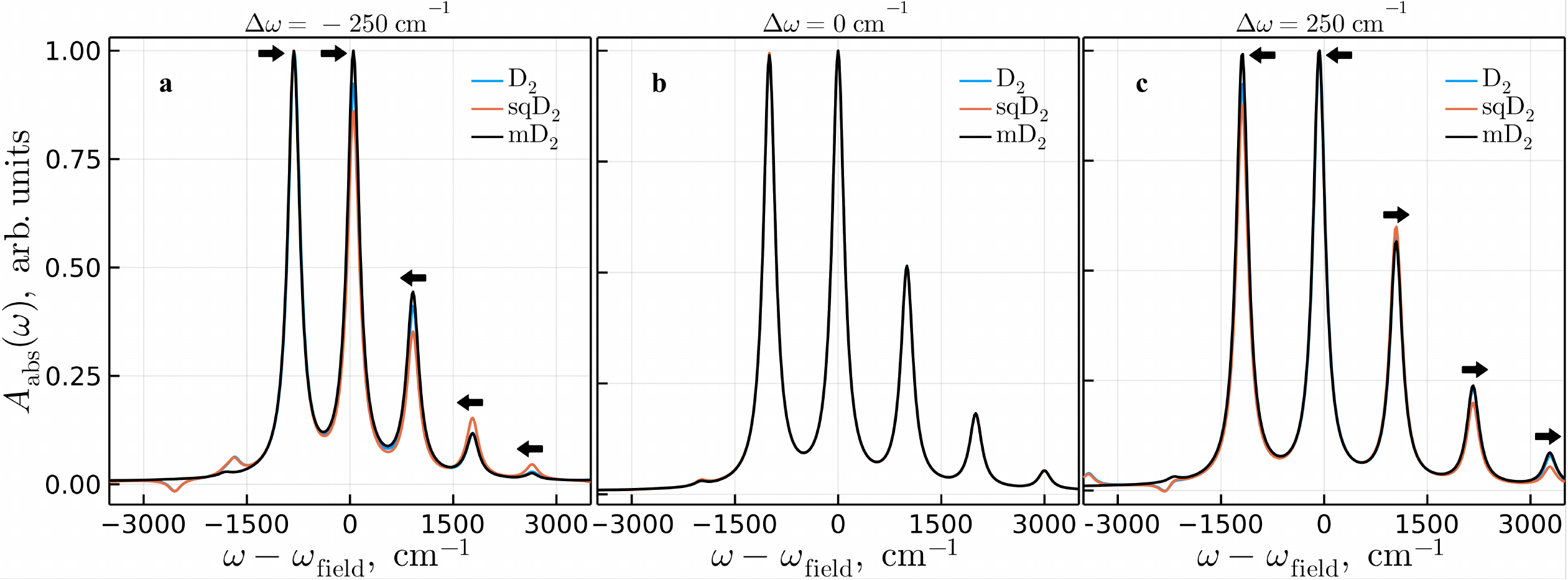}

\caption{Absorption spectrum of a monomer ($\mathcal{M}_{1}$ model) at 300
K temperature with $\Delta\omega_{1,1}$ equal to a) $-250\ \protect\cm$,
b) $0\ \protect\cm$, c) $+250$ $\protect\cm$ simulated using $\text{D}_{2},$
$\text{sqD}_{2}$ and $\text{mD}_{2}$ ansatze. Arrows indicate peak
frequency shift direction when compared to the $\Delta\omega=0$ case.\label{fig:abs_n1}}
\end{figure*}

We assume spontaneus emission to only occur from the lowest energy
excited aggregate vibronic state. After initial excitation by an external
field, due to non-radiative relaxation processes and interaction with
an environment, during the sufficiently long waiting time, $\tau\to\infty$,
aggregate relaxes towards the minimal energy $E_{0}^{(\text{e})}$
excited aggregate vibronic state, $|\Psi_{\text{E}_{0}}^{\left(\text{e}\right)}\left(\tau\right)\rangle$.
From TRF response function in Eq. (\ref{eq:trf_full}) now follows
that the fluorescence response function can be written as 
\begin{equation}
S_{\text{flor}}^{\left(1\right)}\left(t\right)=\langle\Psi_{\text{E}_{\text{0}}}^{\left(\text{e}\right)}\left(0\right)|e^{-i\hat{H}t}\hat{\mu}_{-}e^{-i\hat{H}_{\text{G}}t}\hat{\mu}_{-}|\Psi_{\text{E}_{0}}^{\left(\text{e}\right)}\left(0\right)\rangle,\label{eq:flor}
\end{equation}
where , for convenience, we set the long waiting time to $\tau=0$.
Note, that Eq. (\ref{eq:flor}) does not contain summation over thermal
ensemble trajectories $\gamma$, as the the minimal energy $E_{0}^{(\text{e})}$
and initial state $|\Psi_{\text{E}_{0}}^{\left(\text{e}\right)}\left(0\right)\rangle$
does not depend on initial vibrational conditions (temperature), but
is solely a function of Hamiltonian and chosen ansatz.

The lowest energy state $|\Psi_{E_{0}}^{\left(\text{e}\right)}\left(0\right)\rangle$
is obtained by numerical optimization of excited state energy. That
is obtained using heuristic adaptive particle swarm optimization algorithm
\citep{Zhan2009,mogensen2018optim} by minimizing the total aggregate
energy $E=\langle\Psi|\hat{H}|\Psi\rangle$, as a function of respective
ansatz free parameters. For a given model of interest, optimization
has to be performed once and can be reused afterwards.

At finite temperature $T$, due to thermal energy fluctuations, the
resulting thermal ensemble in the excited aggregate state has larger
average energy $\left\langle E_{0}^{(\text{e})}\right\rangle _{T}\geq E_{0}^{(\text{e})}$.
Therefore, after waiting time $\tau$ aggregate can be in any one
of the thermal ensemble states. Now fluorescence response function
$S_{\text{flor}}^{\left(1\right)}\left(t\right)$ is obtained by averaging
over an ensemble of thermal excited states $|\Psi_{\text{E}_{T}}^{\left(\text{e}\right)}\left(0\right)\rangle_{\gamma}$,
where $\gamma$ is a trajectory number.

In order to find $|\Psi_{\text{E}_{T}}^{\left(\text{e}\right)}\left(0\right)\rangle_{\gamma}$
states, we cannot use the same algorithm as for the electronic ground
state since all vibrational modes in electronic excited state are
now indirectly coupled. Additionally, their frequencies are shifted
if the quadratic vibronic coupling contributes.

For each trajectory $\gamma$, thermal excited states $|\Psi_{\text{E}_{T}}^{\left(\text{e}\right)}\left(0\right)\rangle_{\gamma}$
is obtained by perturbing $|\Psi_{\text{E}_{0}}^{\left(\text{e}\right)}\left(0\right)\rangle$
free parameters in such a way as to increase its total energy by the
energy fluctuation $\delta E_{\gamma}=\sum_{n,q}\omega_{nq}^{\left(\text{e}\right)}\left|\tilde{\lambda}_{nq}^{\left(\gamma\right)}\right|^{2}$,
where $\tilde{\lambda}_{nq}^{\left(\gamma\right)}$ are sampled from
the excited state $\mathcal{P}^{\left(\text{e}\right)}$ distribution
in Eq. (\ref{eq:Sudarshan-P-1}). In order to find free parameters
that correspond to energy $E_{\gamma}=E_{0}^{(\text{e})}+\delta E_{\gamma}$,
we perturb CS displacements $\lambda_{nq}\left(\tau\right)$ for $\text{D}_{2}$,
$\text{sqD}_{2}$ ansatze and $\lambda_{i,nq}\left(\tau\right)$ for
$\text{mD}_{2}$ ansatz, until the new state energy $E_{\gamma}^{\text{fit}}=\langle\Psi_{\text{E}_{T}}^{\left(\text{e}\right)}\left(0\right)|_{\gamma}\hat{H}|\Psi_{\text{E}_{T}}^{\left(\text{e}\right)}\left(0\right)\rangle_{\gamma}$
matches $E_{\gamma}$ with $0.1\ \cm$ precision. Fluorescence response
function at finite temperature is then equal to
\begin{equation}
S_{\text{flor}}^{\left(1\right)}\left(t\right)=\frac{1}{\Gamma}\sum_{\gamma=1}^{\Gamma}\langle\Psi_{\text{E}_{T}}^{\left(\text{e}\right)}\left(0\right)|_{\gamma}e^{-i\hat{H}t}\hat{\mu}_{-}e^{-i\hat{H}_{\text{G}}t}\hat{\mu}_{-}|\Psi_{\text{E}_{T}}^{\left(\text{e}\right)}\left(0\right)\rangle_{\gamma}.
\end{equation}

\section{Results\label{sec:Results}}

\subsection{Model parameters}

In this section we investigate effects of intermolecular coupling
and vibrational mode frequency shifts in Eq. (\ref{eq:H-nonlinear}),
on absorption and fluorescence spectra. We consider three models.
First model, $\mathcal{M}_{1}$, contains a single pigment coupled
to one high frequency intramolecular mode. Second, $\mathcal{M}_{2}$,
is a J-type dimer of two coupled chromophores, where excitations are
coupled to a single high frequency intramolecular vibrational mode
(one per pigment). Third, $\mathcal{M}_{3}$, is again two chromophore
system, but here electronic excitations are coupled to overdamped
phonon bath.
\begin{figure*}
\includegraphics[width=0.8\textwidth]{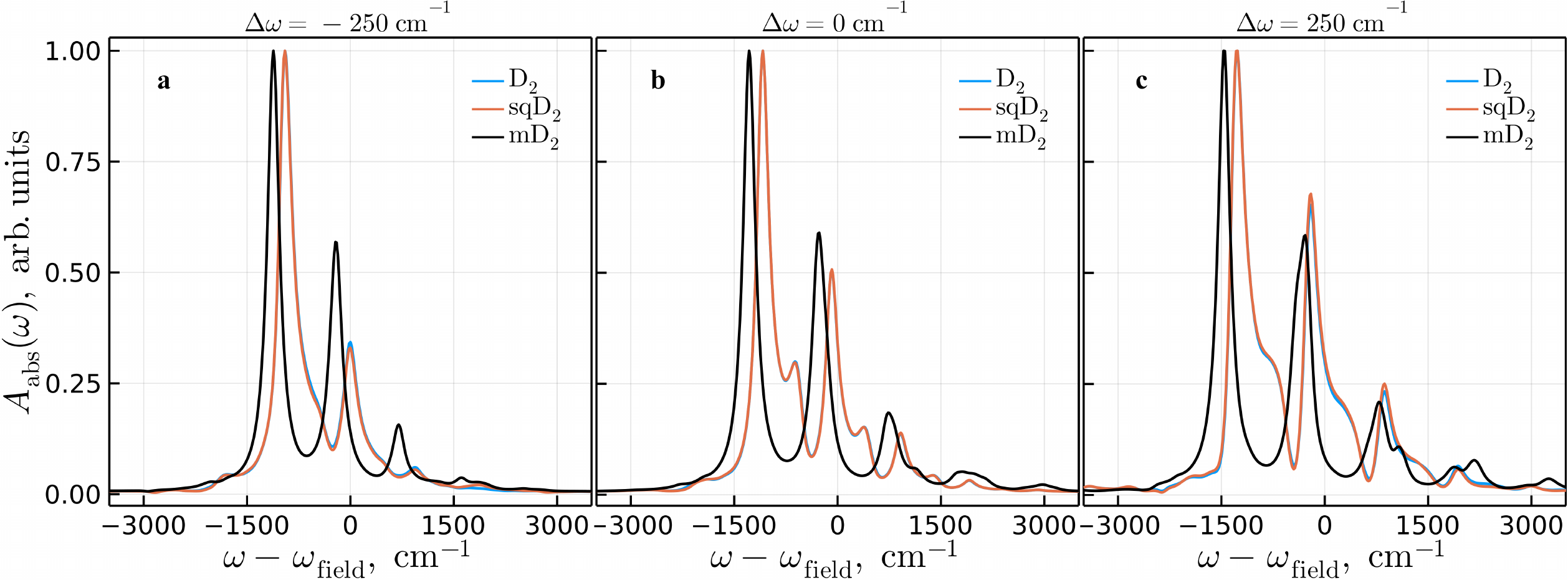}

\caption{Absorption spectrum of a J-type dimer ($\mathcal{M}_{2}$ model) at
300 K temperature with $\Delta\omega_{1,1}=\Delta\omega_{2,1,}$ equal
to a) $-250\ \protect\cm$, b) $0\ \protect\cm$, c) $+250$ $\protect\cm$
simulated using $\text{D}_{2},$ $\text{sqD}_{2}$ and $\text{mD}_{2}$
ansatze. \label{fig:abs_n2}}
\end{figure*}

The J-type dimers in models $\mathcal{M}_{2}$ and $\mathcal{M}_{3}$
consist of two pigments, each of which can be resonantly excited by
an external electric field, thus we assume that single pigment excitation
energies are resonant with optical field, $\varepsilon_{n}=\omega_{\text{field}}$,
where $\omega_{\text{field}}$ is an external field frequency. Electronic
transition dipole moment vectors of the chromophores are identical,
$\boldsymbol{\mu}_{n}=\left(1,0,0\right)$, in Cartesian coordinate
system. For $\mathcal{M}_{1}$ model, intramolecular vibrational mode
frequency in the electronic ground state is $\omega_{1,1}^{g}=1000\ \cm$
and Huang-Rhys (HR) factor is $S=f_{1,1}^{2}=1$. For $\mathcal{M}_{2}$
model, the resonance coupling is $J_{12}=-500\ \cm$, while vibrational
mode frequencies of the chromophores are $\omega_{1,1}^{g}=\omega_{2,1}^{g}=1000\ \cm$
with HR factors $S=f_{1,1}^{2}=f_{2,1}^{2}=1$. For $\mathcal{M}_{3}$
model, the resonance coupling is $J_{12}=-50\ \cm$ and vibrational
phonon mode frequencies $\omega_{n,q}^{g}$ span from $0.1\ \cm$
to $490.1\ \cm$ with step-size of $10\ \cm$ for each pigment $n$
to represent overdamped phonon bath with a given spectral density.
Here the $f_{nq}$ distribution is defined in terms of discretized
quasi-continuos spectral density function
\begin{equation}
C_{n}^{"}\left(\omega\right)=\pi\sum_{q}f_{nq}^{2}\omega_{nq}^{e}\delta(\omega-\omega_{n,q}),
\end{equation}
where $C_{n}^{"}\left(\omega\right)=\omega/\left(\omega^{2}+\gamma^{2}\right)$
is the Drude function with damping $\gamma=100\ \cm$. Magnitudes
of $f_{nq}$ are then normalized so that the total reorganization
energy $\Lambda_{n}=100\ \cm$ for each pigment $n$.

Models $\mathcal{M}_{1}$ and $\mathcal{M}_{2}$ are typically found
in synthetic pigment aggregates \citep{Lim2015a,Christensson2011,Bondarenko2020},
while $\mathcal{M}_{3}$ model more closely corresponds to chlorophyl
aggregates found in nature \citep{Amerongen2010,Valkunas2013a}.

When plotting the simulated absorption and fluorescence response functions
according to the Eq. (\ref{eq:Abs}), we will include phenomenological
dephasing rate of $\gamma_{\text{dep}}=50\ \text{fs}$ for models
$\mathcal{M}_{1}$, $\mathcal{M}_{2}$ and rate of $\gamma_{\text{dep}}=250\ \text{fs}$
for model $\mathcal{M}_{3}$. These are to account for additional
dephasing stemming from explicitly not included phonons (for models
$\mathcal{M}_{1}$, $\mathcal{M}_{2}$) and chromophone vibrational
modes (for model $\mathcal{M}_{3}$).

\subsection{Absorption spectra}

In all models, we vary vibrational mode frequencies in the excited
state $\omega_{nq}^{e}$ by shifting them from frequencies in the
ground state $\omega_{nq}^{g}$, thus we define the difference of
frequencies as $\Delta\omega_{nq}=\omega_{nq}^{e}-\omega_{nq}^{g}$.
First, we start by investigating absorption spectrum of $\mathcal{M}_{1}$
model. In Fig. (\ref{fig:abs_n1}) we present absorption spectrum
of the monomer at 300 K temperature with frequency shifts of $\Delta\omega_{1,1}\equiv\Delta\omega=-250,\ 0,\ +250\ \cm$.
.

When $\Delta\omega=0$, we observe absorption spectrum with vibrational
peak progresion representing jumps from ground to an arbitrary vibrational
excited state. All three ansatze produce identical spectra since there
is no electronic coupling and the nonlinear effects, due to quadratic
vibronic coupling, are also absent. Now, when vibrational mode frequency
in the excited state is higher than the ground state ($\Delta\omega=250\ \cm$),
nonlinear effects become evident together with non-physical features
in spectra of some ansatze. Absorption spectra of $\text{D}_{2}$
and $\text{sqD}_{2}$ ansatz have a negative peak at $\approx2000\ \cm$
suggesting that they are unable to fully capture the nonlinear effects,
i.e., they are not exact solutions of the Schrödinger equation. Meanwhile,
$\text{mD}_{2}$ ansatz with $M=5$ superposition terms produce strictly
positive absorption spectra and thus will be considered to be the
reference spectra for further comparisons. To check validity of this
claim, we compared $\text{mD}_{2}$ spectra simulated with $M=1-10$
terms and found $M\geq5$ spectra to be quantitatively exact (not
shown). Besides the negative peaks, neither $\text{D}_{2}$ nor $\text{sqD}_{2}$
are able to reproduce vibrational peak progression amplitudes of $\text{mD}_{2}$
ansatz.
\begin{figure*}
\includegraphics[width=0.8\textwidth]{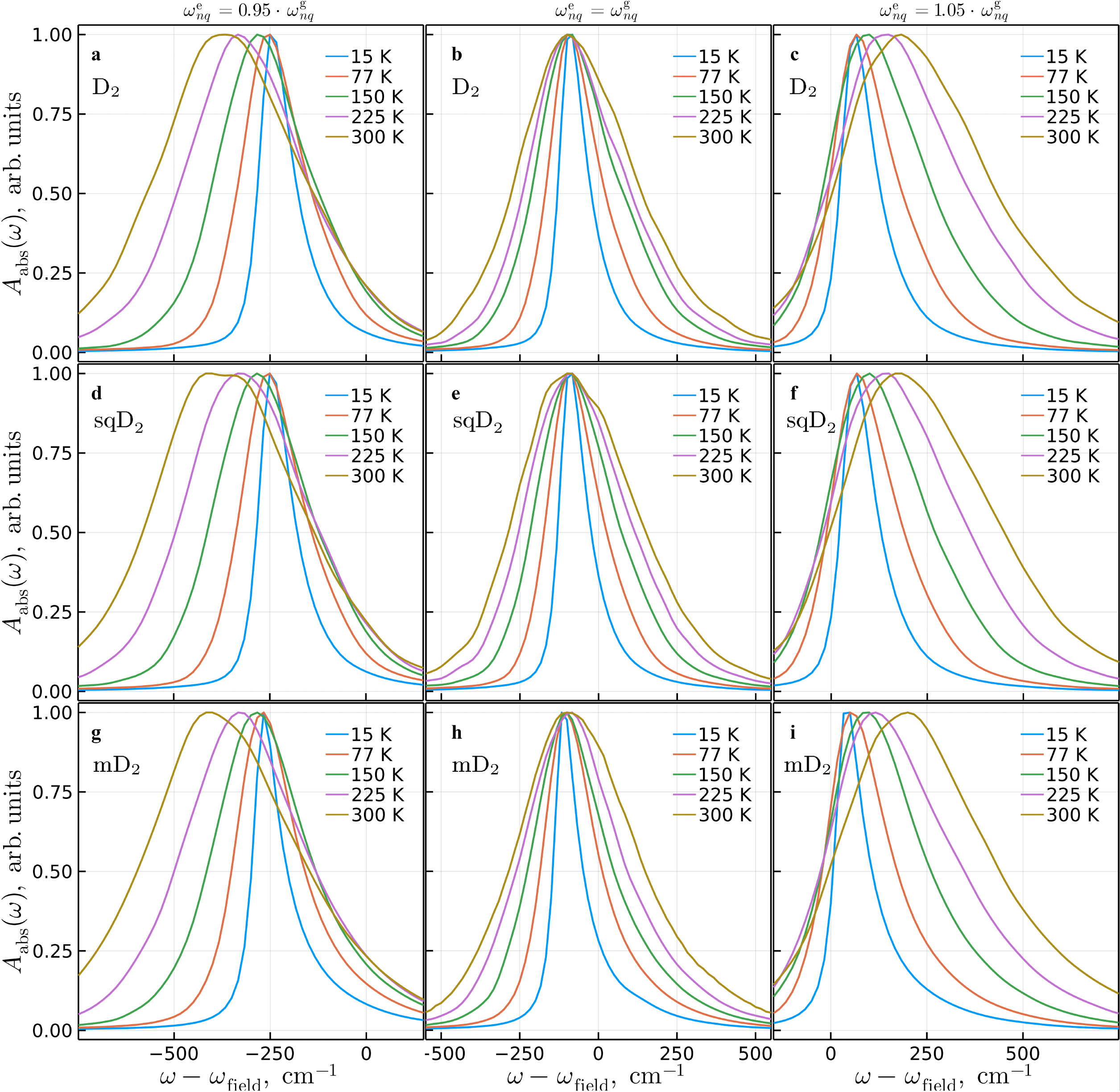}

\caption{Absorption spectrum of a J dimer ($\mathcal{M}_{3}$ model) coupled
to the phonon bath at various temperatures with vibrational mode frequencies
in aggregate excited states $\omega_{nq}^{e}$ equal to a) $0.95\cdot\omega_{nq}^{g}$,
b) $\omega_{nq}^{g}$, c) $1.05\cdot\omega_{nq}^{g}$ simulated using
$\text{D}_{2}$ (1st row), $\text{sqD}_{2}$ (2nd row) and $\text{mD}_{2}$
(3rd row) ansatze. \label{fig:abs_n2_bath}}
\end{figure*}

By comparing $\text{mD}_{2}$ absorption spectra peak amplitude progression
in all three $\Delta\omega$ cases, we find that progression peak
amplitudes either increase or are reduced as compared to the $\Delta\omega=0$
spectrum, we will refer to these qualitative changes as model having
increased or decreased effective HR factor. Therefore, effective HR
is reduced when $\Delta\omega$ is positive, and is increased when
$\Delta\omega$ is negative. In addition, $\Delta\omega$ also changes
progression peak frequencies, however, not in a monotonic fashion.
Direction of frequency change of each peak is indicated by an arrow,
when compared to the $\Delta\omega=0$ case. Absolute frequency of
some peaks increase, while for others it decreases. This can also
be interpreted as relative energy gap between progression peaks becoming
larger when $\Delta\omega$ is positive, and gap is reduced when $\Delta\omega$
is negative.

Next, we look at absorption spectrum of model $\mathcal{M}_{2}$.
In Fig. (\ref{fig:abs_n2}) we present absorption spectrum of the
J dimer at 300 K temperature with shifts $\Delta\omega_{1,1}=\Delta\omega_{2,1,}\equiv\Delta\omega=-250,\ 0,\ 250\ \cm$.
Now, even in the $\Delta\omega=0$ case, when nonlinear effects are
still absent, we find mismatch between absorption spectrum simulated
using $\text{D}_{2},\text{sqD}_{2}$ ansatze and $\text{mD}_{2}$.
This is purelly due to electronic coupling between vibronic states
of sites, which was lacking in model $\mathcal{M}_{1}$. Also, notice
that spectra of $\text{D}_{2}$, $\text{sqD}_{2}$ ansatze are identical,
since according equations $\text{sqD}_{2}$ becomes different from
$\text{D}_{2}$ only when quadratic vibronic coupling is present,
i.e. $\Delta\omega\neq0$. The exact $\text{mD}_{2}$ spectrum has
a familiar J dimer absorption lineshape dominated by the exchange
narrowing effect \citep{Hestand2018}, which effectively reduces HR
factor as compared to the monomer in Fig. (\ref{fig:abs_n1}). Absorption
spectra of $\text{D}_{2}$, $\text{sqD}_{2}$ ansatze reproduces exchange
narrowing effect, however, their spectra has additional secondary
peaks not seen in $\text{mD}_{2}$ spectrum. Their spectra also has
slightly higher energy 0-0 quanta transitions peak (and 0-1, 0-2,
etc.) as compared to the $\text{mD}_{2}$ spectrum, which implies
that $\text{mD}_{2}$ is able to better represent lower energy excited
aggregate state.

When the quadratic vibronic coupling effects are present ($\Delta\omega=-250,\ 250\ \cm$),
again, in both cases, we find $\text{D}_{2},\text{sqD}_{2}$ spectra
to differ from $\text{mD}_{2}$ spectra. Very slight differences can
also be seen between $\text{D}_{2}$ and $\text{sqD}_{2}$ ansatze,
however, without any obvious improvement from $\text{sqD}_{2}$. In
both cases, $\text{mD}_{2}$ spectra again shows J dimer exchange
narrowing type lineshape with changes to peak amplitudes similar to
those seen in Fig. (\ref{fig:abs_n1}) -- relative energy gap between
peaks become larger when $\Delta\omega$ is positive, and is reduced
when $\Delta\omega$ is negative. Spectrum with $\Delta\omega=250\ \cm$
has a more pronounced fine structure to its absorption progression
peaks than those spectra with $\Delta\omega=-250\ \cm$ and $=0$.

These findings suggest that neither $\text{D}_{2}$ nor a more complicated
$\text{sqD}_{2}$ are able to fully capture absorption spectrum of
J dimers with high frequency intramolecular vibrational modes, not
even in the simplest case ($\Delta\omega=0\ \cm$) when the quadratic
vibronic coupling is excluded..

Next, lets look at the absorption spectra of $\mathcal{M}_{3}$ model.
In this case phonon modes become thermally excited so we additionally
present temperature-dependent spectra. In Fig. (\ref{fig:abs_n2_bath})
we present absorption spectra of a J dimer coupled to phonon bath
at various temperatures. Now each chromophone couples to 50 low frequency
vibrational modes, therefore, to investigate quadratic vibronic coupling
effect, we will look at cases when all modes' frequencies, $\omega_{nq}^{e}$,
in an excited aggregate are equal to frequencies in an ground aggregate
state, $\omega_{nq}^{g}$, scaled by a factor of $\gamma=0.95,\ 1,\ 1.05$.
\begin{figure*}
\includegraphics[width=0.8\textwidth]{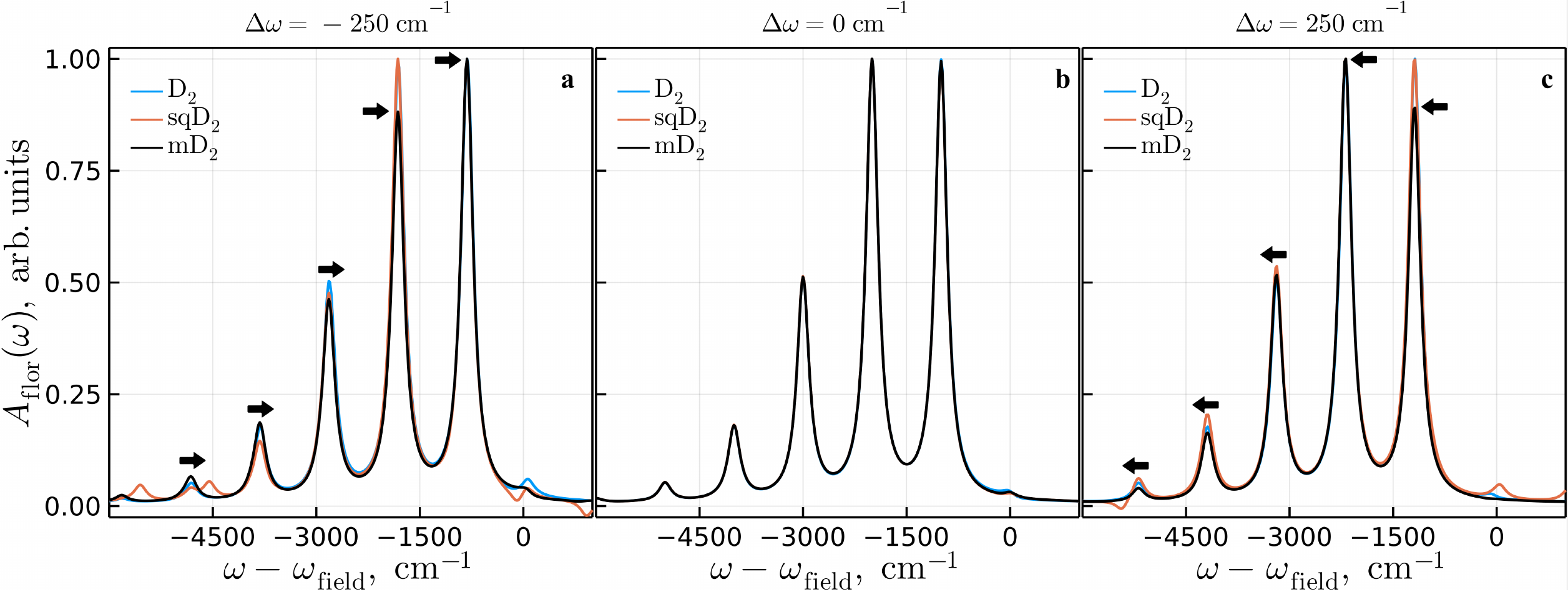}

\caption{\label{fig:flor_n1} Fluorescence spectrum of a monomer ($\mathcal{M}_{1}$
model) at 300 K temperature with $\Delta\omega_{1,1}$ equal to a)
$-250\ \protect\cm$, b) $0\ \protect\cm$, c) $+250$ $\protect\cm$
simulated using $\text{D}_{2},$ $\text{sqD}_{2}$ and $\text{mD}_{2}$
ansatze. Arrows indicate peak frequency shift direction when compared
to the $\Delta\omega=0$ case.}
\end{figure*}

When $\gamma=1$, all methods produce qualitatively identical absorption
spectra over a broad range of temperatures. At low temperatures spectra
consists of a single absorption peak. With increasing temperature,
spectra broadens and slightly shifts (on average) due to thermal excitation
of vibrational modes in electronic ground state and due to finite
discretization at low frequencies. .

Now, when phonon mode frequencies in the aggregate excited state are
higher ($\gamma=1.05$), in addition to the previously seen thermal
spectra broadening, we also observe two type of spectral shifts: a
static shift -- the whole absorption spectra shifts to the higher
energies, as compared to the $\gamma=1$ case, and a temperature dependent
absorption peak shift to the higher energies. Spectra simulated with
all ansatze when $\gamma=1.05$ are also qualitativelly similar, however,
spectrum of $\text{mD}_{2}$ in Fig. (\ref{fig:abs_n2_bath}i) has
a less straightforward temperature dependent peak shift dependence.
Peak frequency changes not as linearly with temperature as in spectra
simulated with $\text{D}_{2}$ and $\text{sqD}_{2}$ ansatze in Fig.
(\ref{fig:abs_n2_bath}c) and Fig. (\ref{fig:abs_n2_bath}f). Similarly,
when phonon mode frequencies in the aggregate excited state are lower
($\gamma=0.95$), we find all the same spectral shift effects, only
now to the lower energy side.Spectra simulated with different ansatze
appear qualitatively the same, therefore we conclude that to simulate
absorption spectra of J dimer coupled to low frequency phonon modes,
even with quadratic vibronic coupling, it is sufficient to use the
simplest $\text{D}_{2}$ ansatz.Fluorescence spectra

\begin{table}
\begin{centering}
\begin{tabular}{|c|c|c|c|c|}
\hline 
\multirow{4}{*}{$\mathcal{M}_{1}$} & $\Delta\omega$ & $\text{D}_{2}$ & $\text{sqD}_{2}$ & $\text{mD}_{2}$\tabularnewline
\cline{2-5} \cline{3-5} \cline{4-5} \cline{5-5} 
 & $-250$ & -812.5 & -812.5 & -816.9\tabularnewline
\cline{2-5} \cline{3-5} \cline{4-5} \cline{5-5} 
 & 0 & -1000.0 & -1000.0 & -1000.0\tabularnewline
\cline{2-5} \cline{3-5} \cline{4-5} \cline{5-5} 
 & $250$ & -1187.5 & -1190.9 & -1190.9\tabularnewline
\hline 
\multicolumn{1}{c}{} & \multicolumn{1}{c}{} & \multicolumn{1}{c}{} & \multicolumn{1}{c}{} & \multicolumn{1}{c}{}\tabularnewline
\hline 
\multirow{4}{*}{$\mathcal{M}_{2}$} & $\Delta\omega$ & $\text{D}_{2}$ & $\text{sqD}_{2}$ & $\text{mD}_{2}$\tabularnewline
\cline{2-5} \cline{3-5} \cline{4-5} \cline{5-5} 
 & $-250$ & -956.3 & -959.9 & -1119.2\tabularnewline
\cline{2-5} \cline{3-5} \cline{4-5} \cline{5-5} 
 & $0$ & -1125.0 & -1125.0 & -1284.7\tabularnewline
\cline{2-5} \cline{3-5} \cline{4-5} \cline{5-5} 
 & $250$ & -1131.9 & -1302.0 & -1460.9\tabularnewline
\hline 
\multicolumn{1}{c}{} & \multicolumn{1}{c}{} & \multicolumn{1}{c}{} & \multicolumn{1}{c}{} & \multicolumn{1}{c}{}\tabularnewline
\hline 
\multirow{4}{*}{$\mathcal{M}_{3}$} & $\omega_{nq}^{e}$ & $\text{D}_{2}$ & $\text{sqD}_{2}$ & $\text{mD}_{2}$\tabularnewline
\cline{2-5} \cline{3-5} \cline{4-5} \cline{5-5} 
 & $0.95\cdot\omega_{nq}^{g}$ & -265.09 & -265.1 & -265.0\tabularnewline
\cline{2-5} \cline{3-5} \cline{4-5} \cline{5-5} 
 & $1.0\cdot\omega_{nq}^{g}$ & -112.5 & -112.2 & -111.7\tabularnewline
\cline{2-5} \cline{3-5} \cline{4-5} \cline{5-5} 
 & $1.05\cdot\omega_{nq}^{g}$ & 41.1 & 40.4 & 41.1\tabularnewline
\hline 
\end{tabular}
\par\end{centering}
\caption{Energy $E_{0}^{(\text{e})}$ of aggregate excited state $|\Psi_{\text{E}_{0}}^{\left(\text{e}\right)}\left(0\right)\rangle$
for models $\mathcal{M}$ using $\text{D}_{2}$, $\text{sqD}_{2}$
and $\text{mD}_{2}$ ansatze. $\text{mD}_{2}$ ansatz consist of $M=5$
superposition terms. Values are in units of $\protect\cm$.\label{tab:energies}}
\end{table}

In order to compute fluorescence spectrum, for each considered ansatze,
we first have to find the lowest energy $E_{0}^{(\text{e})}$ excited
aggregate state $|\Psi_{E_{0}}^{\left(\text{e}\right)}\left(0\right)\rangle$
in terms of that ansatz free parameter by minimizing the total aggregate
energy $E=\langle\Psi|\hat{H}|\Psi\rangle$, as explained in Section
(\ref{subsec:Absorption-and-fluorescence theory}). The resulting
energies $E_{0}^{(\text{e})}$ for models $\mathcal{M}$ are given
in Table (\ref{tab:energies}).
\begin{figure*}
\includegraphics[width=0.8\textwidth]{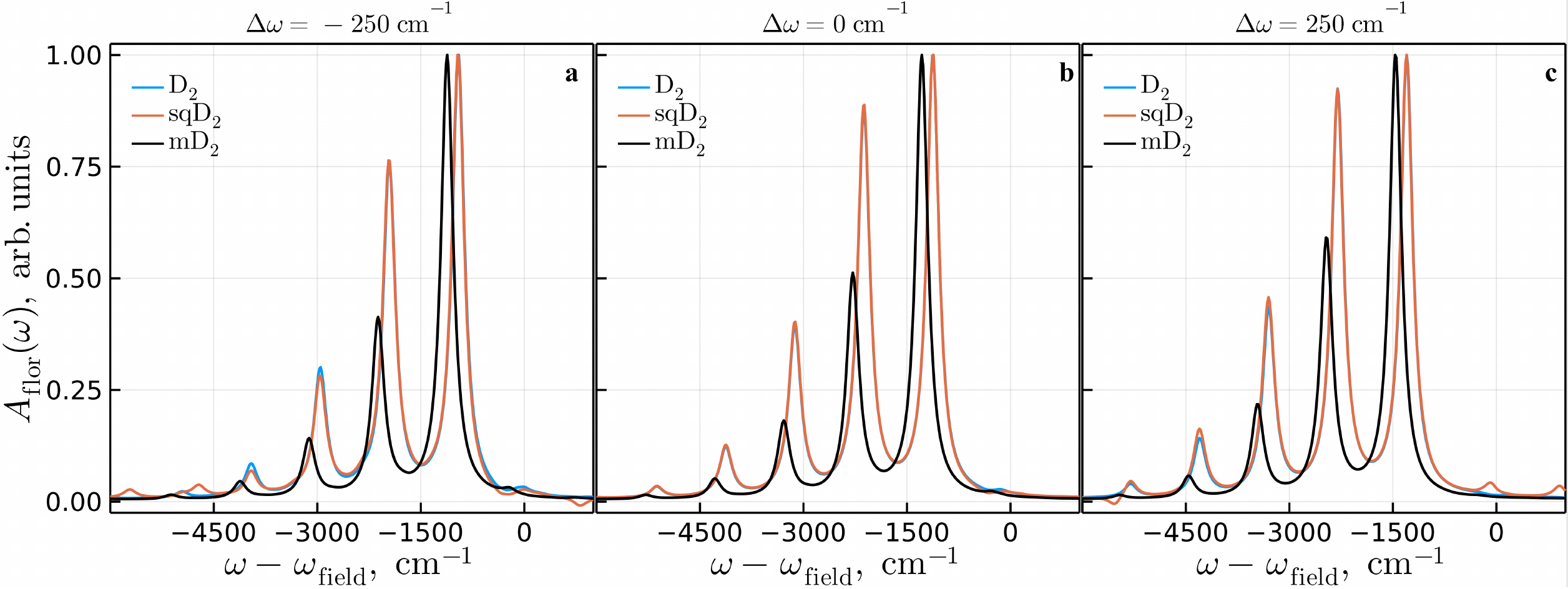}

\caption{\label{fig:flor_n2} Fluorescence spectrum of a J dimer ($\mathcal{M}_{2}$
model) at 300 K temperature with $\Delta\omega_{1,1}=\Delta\omega_{2,1,}$
equal to a) $-250\ \protect\cm$, b) $0\ \protect\cm$, c) $250$
$\protect\cm$ simulated using $\text{D}_{2},$ $\text{sqD}_{2}$
and $\text{mD}_{2}$ ansatze.}
\end{figure*}

We see that for model $\mathcal{M}_{1}$, when vibrational nonlinearities
are absent, all ansatze give exactly the same energy, however, by
including the quadratic vibronic coupling ($\Delta\omega\neq0$ cases),
both $\text{sqD}_{2}$ and $\text{mD}_{2}$ find lower energy states
than $\text{D}_{2}$ ansatz. $\text{mD}_{2}$ ansatz further outperform
$\text{sqD}_{2}$ ansatz, when $\Delta\omega$ is negative. Consequently,
$\text{sqD}_{2}$ model outperforms $\text{D}_{2}$ ansatz when searching
for excited state energy minimum when quadratic coupling is included.

In model $\mathcal{M}_{2}$, when $\Delta\omega=0$, we see that $\text{D}_{2}$
and $\text{sqD}_{2}$ again find equivalent energy state, however,
now $\text{mD}_{2}$ ansatz manages to represent significantly lower
energy state, which is not accessed by any of the non-multiple ansatze
and is created purelly due to electronic coupling between pigments.
When nonlinearities are included, $\text{sqD}_{2}$ ansatz again outperforms
$\text{D}_{2}$, especially when $\Delta\omega$ is positive, yet
$\text{mD}_{2}$ further improves on $\text{sqD}_{2}$ states.

In model $\mathcal{M}_{3}$, we try to find minimum point in 1020-dimensional
space for $\text{mD}_{2}$, 204-dimensional for $\text{D}_{2}$, and
404-dimensional for $\text{sqD}_{2}$, which is a difficult problem
to solve. To have a fair comparison of ansatze for model $\mathcal{M}_{3}$,
we limited search for the state $|\Psi_{E_{0}}^{\left(\text{e}\right)}\left(0\right)\rangle$
in terms of $\text{sqD}_{2}$ ansatz to its sqCS displacement parameters,
$\lambda_{kq}$, and set squeezing parameters to $r_{kq}=1$, $\theta_{kq}=0$
(no squeezing). For the $\text{mD}_{2}$ ansatz, we limited search
to just one of its multiples. With these limits set, essentially both
$\text{sqD}_{2}$ and $\text{mD}_{2}$ ansatz behave as $\text{D}_{2}$,
thus all three ansatze relax to the same excited aggregate state with
energies equivalent to those under $\text{D}_{2}$ column. This is
confirmed by numberical results where all ansatze managed to represent
states with very similar energies. The obtained numbers are also likely
within the margin of error and require an improved approach for finding
actual lowest energy states.

Now, lets look at fluorescence spectra of the same $\mathcal{M}$
models. In Fig. (\ref{fig:flor_n1}) we display fluorescence spectra
of a monomer coupled to high frequency vibration ($\mathcal{M}_{1}$
model) at 300 K temperature with frequency shifts of $\Delta\omega_{1,1}\equiv\Delta\omega=-250,\ 0,\ +250\ \cm$.
When $\Delta\omega=0$, we find all three ansatze to produce identical
fluorescence spectra, which, as expected, has a mirror symmetry with
$\mathcal{M}_{1}$ model absorption spectrum in Fig. (\ref{fig:abs_n1}b).
Fluorescence spectrum consists of progression of energeticaly downward
transition peaks.

When the quadratic vibronic coupling term is included ($\Delta\omega=-250,\ +250\ \cm$),
simulated fluorescence spectra of considered ansatze are different.
In both cases, fluorescence spectrum of $\text{D}_{2}$ qualitatively
match $\text{mD}_{2}$ ansatz spectrum peak amplitudes and frequencies,
while the intermediate complexity $\text{sqD}_{2}$ consistently overestimate
peak amplitudes and show additional peaks that are not present in
$\text{mD}_{2}$ ansatz spectrum. Here we see an example, where additional,
but not sufficient, DOF (squeezing) of $\text{sqD}_{2}$ ansatz actually
produce visually worse quality spectrum than the smaller state space
$\text{D}_{2}$ ansatz. This is in contrast to absorption spectra
of $\mathcal{M}_{1}$ model, where both $\text{D}_{2}$ and $\text{sqD}_{2}$
ansatz showed equivalent errors when compared to $\text{mD}_{2}$
spectra.

By comparing $\text{mD}_{2}$ fluorescence spectra with quadratic
vibronic coupling to that without it, we find that fluorescence progression
peak amplitudes change -- effective HR factor increases when $\Delta\omega$
is positive, and decreases when $\Delta\omega$ is negative. Also,
quadratic vibronic coupling shifts whole spectra to the lower energy
side when $\Delta\omega$ is positive, and to the higher side when
$\Delta\omega$ is negative. In contrast to the absorption spectra
in Fig. (\ref{fig:abs_n1}), energy gaps between progression peaks
remain unchanged, Also, by comparing quadratic vibronic coupling absorption
and fluoresnce spectra of $\mathcal{M}_{1}$ model of $\text{mD}_{2}$
ansatz, we see that quadratic vibronic coupling breaks the mirror
symmetry between the two.

Now, lets move on to the $\mathcal{M}_{2}$ model. In Fig. (\ref{fig:flor_n2})
we show its fluorescence spectrum simulated at 300 K temperature with
$\Delta\omega_{1,1}=\Delta\omega_{2,1,}\equiv\Delta\omega$ equal
to $-250\ \cm$, $0\ \cm$, $250$ $\cm$.

When nonlinearities are absent ($\Delta\omega=0$), we again see that
$\text{D}_{2}$ and $\text{sqD}_{2}$ ansatze yeld identical fluorescence
spectra, which differ from the spectrum of $\text{mD}_{2}$ ansatz
in fluorescence peak amplitudes and frequencies. The discrepency between
spectra is again a result of $\text{D}_{2}$, $\text{sqD}_{2}$ ansatze
not being able to properly represent vibronic states created by electronic
coupling between J dimer pigments. The lineshape of $\text{mD}_{2}$
fluorescence spectrum is dominated by the exchange narrowing effect
and does not have mirror symmetry with absorption spectrum.

From fluorescence spectra of $\mathcal{M}_{2}$ model with the quadratic
vibronic coupling term ($\Delta\omega=-250,\ +250\ \cm$), we draw
the same conclusions as in the $\mathcal{M}_{1}$ model: $\text{D}_{2}$
spectrum matches $\text{mD}_{2}$ spectrum better than does $\text{sqD}_{2}$;
effective HR factor increases when $\Delta\omega$ is positive, and
decreases when $\Delta\omega$ is negative; quadratic vibronic coupling
shift spectra to the lower energy side when $\Delta\omega$ is positive,
and to the higher side when $\Delta\omega$ is negative; energy gaps
between progression peaks remain unchanged from $\Delta\omega=0$
spectrum.

Overall, fluorescence spectrum of J dimer coupled to the high frequency
vibrational modes is accuratelly captured only by the $\text{mD}_{2}$
ansatz, while $\text{sqD}_{2}$ yield visually slightly worse quality
spectrum than that of $\text{D}_{2}$ ansatz, however, neither are
to match $\text{mD}_{2}$ accuracy. 
\begin{figure*}
\includegraphics[width=0.8\textwidth]{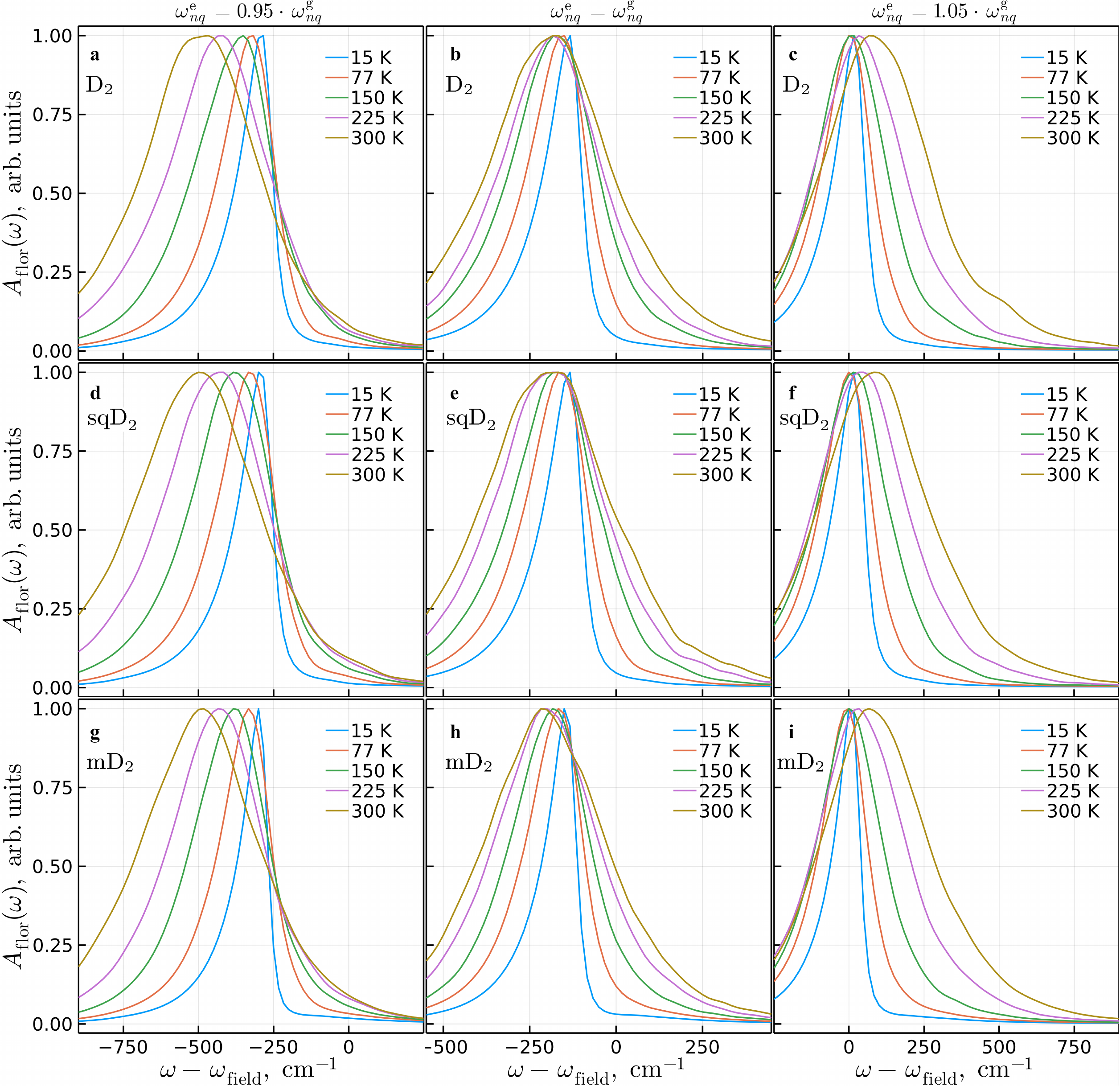}

\caption{\label{fig:flor_bath} Fluorescence spectrum of a J dimer ($\mathcal{M}_{3}$
model) coupled to the bath at various temperatures with vibrational
mode frequencies in aggregate excited states $\omega_{nq}^{e}$ equal
to a) $0.95\cdot\omega_{nq}^{g}$, b) $\omega_{nq}^{g}$, c) $1.05\cdot\omega_{nq}^{g}$
simulated using $\text{D}_{2}$ (1st row), $\text{sqD}_{2}$ (2nd
row) and $\text{mD}_{2}$ (3rd row) ansatze.}
\end{figure*}

Next, lets look at the fluorescence spectra of $\mathcal{M}_{3}$
model. In Fig. (\ref{fig:flor_bath}) we show fluorescence spectra
of a J dimer coupled to bath of low frequency phonon modes at various
temperatures. We see that all ansatze produce qualitatively simillar
fluorescence spectra with all vibrational mode scalling factors $\gamma=0.95,\ 1,\ 1.05$.
As in absorption spectra of $\mathcal{M}_{3}$ model in Fig. (\ref{fig:abs_n2_bath}),
we find analogous effects of spectral broadening with increasing temperature,
as well as two type of spectral shifts: a static shift -- the whole
spectrum shifts to the higher energies when $\gamma$ is positive,
and to the lower energies when $\gamma$ is negative, as compared
to the $\gamma=1$ case, and an additional temperature dependent fluorescence
peak shift to the higher energy side when $\gamma$ is positive, and
to the lower side when $\gamma$ is negative. In addition to these,
we now observe fluorescence peak drift to the lower energies with
increasing temperature when the frequency scale factor is $\gamma=1$,
regardless of the ansatze used.

All in all, spectra simulated with considered ansatze appear qualitatively
equivalent, thus we conclude that to simulate fluorescence spectra
of J dimer coupled to low frequency phonon modes, even with quadratic
vibronic coupling, it is sufficient to use the simplest $\text{D}_{2}$
ansatz.

\section{Discussion}

Natural progression in constructing more and more sophisticated Davydov
type ansatze, would be to write down ansatz as a superposition of
$\text{sqD}_{2}$ ansatze -- the $\text{multi-sqD}_{2}$ ansatz.
This was recently done by Zeng et al. \citep{Zeng2022a}, where they
used it to simulate dynamics and absorption spectra of pyrazine and
the 2-pyridone dimer aggregate, and found a great match with the state-of-the-art
multi-configuration time-dependent Hartree (MCTDH) method results.
In fact, presented approach of using Davydov type ansatze is closely
related to the Gaussian-MCTDH with frozen Gaussians functions for
$\text{D}_{2}$, $\text{mD}_{2}$ ansatze, and $\text{sqD}_{2}$ with
thawed Gaussian functions \citep{Zhao2021,Beck2000a,Worth2003a,Worth2009}.

Our analysis presented in Section (\ref{sec:Results}), show that
using sqCS, instead of regular CS, does not provide any significant
improvement to the simulated absorption and fluorescence spectra of
J dimers, even when the quadratic vibronic coupling is used. Therefore
one has to wonder if an additional numerical effort needed to propagate
$\text{multi-sqD}_{2}$ ansatz is worth, since any arbitrary wavefunction
can be already exactly expanded using $\text{mD}_{2}$ ansatz using
the unity operator expression
\begin{equation}
\hat{I}=\pi^{-1}\iint\text{d\text{Re}\ensuremath{\lambda}}\ \text{d\text{Im}\ensuremath{\lambda}}\ |\lambda\rangle\langle\lambda|.
\end{equation}
It would be interesting to see if the $\text{multi-sqD}_{2}$ ansatz
would require less terms in its superposition than the $\text{mD}_{2}$
ansatz to obtain equivalent spectra. However, this is outside the
topic of this paper.

We looked at the quadratic vibronic coupling effects for low and high
frequency modes. For the high frequency modes, we looked at large
nonlinearities by increasing and decreasing mode frequency by 25\%,
which is much larger than what is observed in molecules \citep{Jakucionis2022}.
This was chosen to investigate limits of all ansatze, however, for
smaller nonlinearities we expect the same conclusion, i.e., that multiple-type
ansatze are required to simulate aggregate spectra. This is because
we considered strong electronic coupling between pigments, which eventually
splits wavepacket into several discrete packets and move quasi-independent
along seperate vibronic state energy surfaces, while the quadratic
vibronic coupling introduces only the secondary effects, which were
not captured by non-multiple ansatze.

For the low frequency modes, we considered small nonlinearities by
changing frequencies by 5\%, more in line with what is observed, with
small electronic coupling between pigments, and found all considered
ansatze to produce qualitativelly identical spectra. This implies
that even when quadratic vibronic coupling is the main source of nonlinearity,
for realistic frequency shifts, sqCS does not provide any significant
improvement. However, it is worth mentioning that $\text{sqD}_{2}$
model outperforms $\text{D}_{2}$ ansatz when searching for excited
state energy minimum when quadratic coupling is included. This improvement
may be important for other types of processes such as charge separation
and internal conversion.

In conclusion, we compared absorption and fluorescence spectra of
vibronic J dimer model with quadratic vibronic coupling simulated
using three increasing sophistication wavefunction ansatze: $\text{D}_{2}$,
$\text{sqD}_{2}$ and $\text{mD}_{2}$. We found that it is necessary
to use $\text{mD}_{2}$ ansatz whenever molecular aggregate electronic
DOFs are coupled to higher frequency intramolecular vibrational modes.
If they are coupled to low frequency phonon bath modes, all three
ansatze produce qualitatively the same spectra. The quadratic vibronic
coupling term manifests itself in both absorption and fluorescence
spectra as a lineshape peak amplitude redistribution, static frequency
shift and an additional shift, which is dependent on the temperature.

\section*{Conflicts of interest}

There are no conflicts of interest to declare.
\begin{acknowledgments}
We thank the Research Council of Lithuania for financial support (grant
No: SMIP-20-47). Computations were performed on resources at the High
Performance Computing Center, \textquoteleft \textquoteleft HPC Sauletekis\textquoteright \textquoteright{}
in Vilnius University Faculty of Physics.
\end{acknowledgments}

\appendix

\section{Time-dependent variational principle\label{sec:EOM}}

Will be using time-dependent Dirac-Frenkel variational principle to
obtain a set of equations of motion of the $\text{D}_{2},\text{sqD}_{2}$
and $\text{mD}_{2}$ ansatze free parameters: $\Gamma_{\text{D}_{2}}=\left\{ \alpha_{n}\left(t\right),\lambda_{kq}\left(t\right)\right\} $,
$\Gamma_{\text{sqD}_{2}}=\left\{ \alpha_{n}\left(t\right),\lambda_{kq}\left(t\right),r_{kq}\left(t\right)\right\} $
and $\Gamma_{\text{mD}_{2}}=\left\{ \alpha_{i,n}\left(t\right),\lambda_{i,kq}\left(t\right)\right\} $.
Solution of the set of equations will result in ansatze time evolution,
such that the deviation from an exact solution of the Schrödinger
equation will be minimized. As a first step, we write down model Lagrangian
in the form of

\begin{align}
\mathcal{L}\left(t\right) & =\frac{\text{i}}{2}\left(\langle\Psi\left(t\right)|\dot{\Psi}\left(t\right)\rangle-\langle\dot{\Psi}\left(t\right)|\Psi\left(t\right)\rangle\right)\nonumber \\
 & -\langle\Psi\left(t\right)|\hat{H}|\Psi\left(t\right)\rangle
\end{align}
where $\dot{x}\left(t\right)$ is the time derivative of $x\left(t\right)$.
For the $\text{sqD}_{2}$ ansatz, Lagrangian can be expressed as (hereafter,
we omit explicitly writing parameter time dependence)

\begin{align}
\mathcal{L}_{\text{sqD}_{2}} & =\frac{\text{i}}{2}\sum_{n}\alpha_{n}^{*}\dot{\alpha}_{n}-\frac{\mathrm{i}}{2}\sum_{m}\alpha_{m}\dot{\alpha}_{m}^{*}\nonumber \\
 & +\frac{\text{i}}{2}\sum_{n,h,q}\left|\alpha_{n}\right|^{2}\left(\dot{\lambda}_{hq}\lambda_{hq}^{*}-\dot{\lambda}_{hq}^{*}\lambda_{hq}+\text{i}\dot{\theta}_{hq}\sinh^{2}\left(r_{hq}\right)\right)\nonumber \\
 & -\sum_{n}\left|\alpha_{n}\right|^{2}\varepsilon_{n}-\sum_{n,m}^{n\neq m}V_{nm}\alpha_{m}\alpha_{n}^{*}\nonumber \\
 & -\sum_{n,m,q}\left|\alpha_{m}\right|^{2}\omega_{nq}^{g}\left(\sinh^{2}\left(r_{nq}\right)+\left|\lambda_{nq}\right|^{2}\right)\nonumber \\
 & +2\sum_{n,q}\left|\alpha_{n}\right|^{2}\omega_{nq}^{e}f_{nq}\text{Re}\lambda_{nq}\nonumber \\
 & -\sum_{n,q}\left|\alpha_{n}\right|^{2}\Delta\omega_{nq}\nonumber \\
\times & \left(\cosh\left(2r_{nq}\right)-\sinh\left(2r_{nq}\right)\cos\left(\theta_{nq}\right)+\left(2\text{Re}\lambda_{nq}\right)^{2}\right)\ ,
\end{align}
and for the $\text{mD}_{2}$, Lagrangian reads
\begin{align}
\mathcal{L}_{\text{mD}_{2}} & =\text{i}\sum_{i,j}\sum_{n}\alpha_{i,n}^{\star}\dot{\alpha}_{j,n}S_{ij}\nonumber \\
 & +\text{i}\sum_{i,j}\sum_{n}\alpha_{i,n}^{\star}\alpha_{j,n}S_{ij}K_{ij}\nonumber \\
 & -\sum_{i,j}\sum_{n}\alpha_{i,n}^{\star}\alpha_{j,n}S_{ij}\varepsilon_{n}-\sum_{i,j}\sum_{n,m}\alpha_{i,n}^{\star}\alpha_{j,m}S_{ij}J_{nm}\nonumber \\
 & -\sum_{ij}\sum_{n}\alpha_{i,n}^{\star}\alpha_{j,n}S_{ij}\sum_{h}\omega_{nh}\lambda_{i,nh}^{\star}\lambda_{j,nh}\nonumber \\
 & +\sum_{ij}\sum_{n}\alpha_{i,n}^{\star}\alpha_{j,n}S_{ij}\sum_{h}\omega_{nh}f_{nh}\left(\tilde{\lambda}_{i,nh}^{\star}+\tilde{\lambda}_{j,nh}\right)\nonumber \\
 & -\sum_{ij}\sum_{n}\alpha_{i,n}^{\star}\alpha_{j,n}S_{ij}\sum_{h}\Delta\omega_{nh}\left(1+\left(\lambda_{i,nh}^{\star}+\lambda_{j,nh}\right)^{2}\right),
\end{align}
where Debay-Waller factor is 
\begin{align}
S_{ij}= & \exp\left\{ \sum_{k,q}\lambda_{i,kq}^{\star}\lambda_{j,kq}-\frac{1}{2}\left(\left|\lambda_{i,kq}\right|^{2}+\left|\lambda_{j,kq}\right|^{2}\right)\right\} ,\label{eq:Debye-Waller}
\end{align}
and 
\begin{equation}
K_{ij}=\sum_{kq}\lambda_{i,kq}^{\star}\dot{\lambda}_{j,kq}-\frac{1}{2}\frac{\text{d}}{\text{dt}}\left|\lambda_{j,kq}\right|^{2}.
\end{equation}

Now, for each Lagrangian $\mathcal{L}_{\beta}$, where $\beta=\text{sqD}_{2},\ \text{mD}_{2}$,
we applying the Euler-Lagrange equation

\begin{equation}
\frac{\text{d}}{\text{dt}}\left(\frac{\partial\mathcal{L}_{\beta}}{\partial\dot{\gamma_{\beta}}^{\star}}\right)-\frac{\partial\mathcal{L}_{\beta}}{\partial\gamma_{\beta}^{\star}}=0,
\end{equation}
to each free parameters $\gamma_{\beta}\in\Gamma_{\beta}$ of ansatz
in order to obtain equation of motion.

For the $\text{sqD}_{2}$ ansatz, this procedure results in a system
of differential equations:

\begin{align}
\dot{\alpha}_{n}= & -\frac{1}{2}\alpha_{n}\sum_{h,q}\left(\dot{\lambda}_{hq}\lambda_{hq}^{*}-\dot{\lambda}_{hq}^{*}\lambda_{hq}\right)\nonumber \\
 & -\frac{\mathrm{i}}{2}\alpha_{n}\sum_{h,q}\dot{\theta}_{hq}\sinh^{2}\left(r_{hq}\right)\nonumber \\
 & -\mathrm{i}\alpha_{n}\varepsilon_{n}-\mathrm{i}\sum_{m}^{n\neq m}V_{m}\alpha_{m}\nonumber \\
 & -\mathrm{i}\alpha_{n}\sum_{m,q}\omega_{mq}^{g}\left(\sinh^{2}\left(r_{mq}\right)+\left|\lambda_{mq}\right|^{2}\right)\nonumber \\
 & +\mathrm{i}2\alpha_{n}\sum_{q}\omega_{nq}^{e}f_{nq}\text{Re}\lambda_{nq}\nonumber \\
 & -\mathrm{i}\alpha_{n}\sum_{q}\Delta\omega_{nq}\cosh\left(2r_{nq}\right)\nonumber \\
 & +\mathrm{i}\alpha_{n}\sum_{q}\Delta\omega_{nq}\sinh\left(2r_{nq}\right)\cos\left(\theta_{nq}\right)+\left(2\text{Re}\lambda_{nq}\right)^{2},
\end{align}
for each index $n$, and

\begin{align}
\dot{\lambda}_{kh} & =-\mathrm{i}\omega_{kh}^{g}\lambda_{kh}+\frac{\text{\ensuremath{\mathrm{i}}}}{\rho}\left|\alpha_{k}\right|^{2}\omega_{kh}^{e}f_{kh}\nonumber \\
 & -4\frac{\text{i}}{\rho}\left|\alpha_{k}\right|^{2}\Delta\omega_{kh}\text{Re}\lambda_{kh},
\end{align}

\begin{equation}
\dot{r}_{kh}=\frac{\text{2}}{\rho}\left|\alpha_{k}\right|^{2}\Delta\omega_{kh}\sin\left(\theta_{kh}\right),
\end{equation}

\begin{align}
\dot{\theta}_{kh} & =-2\omega_{kh}^{g}\nonumber \\
 & -4\frac{\left|\alpha_{k}\right|^{2}}{\rho}\Delta\omega_{kh}\left(1-\coth\left(2r_{kh}\right)\cos\left(\theta_{kh}\right)\right),
\end{align}
for each pair of $\left\{ k,h\right\} $ indeces.

We denote $\rho=\sum_{n}\left|\alpha_{n}\right|^{2}$ as the total
population. Only the last two terms $r_{kh}$, $\theta_{kh}$, which
make up the complex squeezing parameter $\zeta=r_{kh}e^{\text{i\ensuremath{\theta_{kh}}}}$,
depend on $\Delta w$. Now, if we look back at Hamiltonian terms Eq.
(\ref{eq:H-s}-\ref{eq:H-nonlinear}), we see straightaway that only
the quadratic vibronic term depends on $\Delta w$, thus squeezing
is generated only by this term. Otherwise, if $\Delta w=0$, squeezing
amplitude $r_{kh}$ becomes time independent, while squeezing angle
$\theta_{kh}$ changes at a constant rate $-2\omega_{kh}^{g}$.

For the $\text{mD}_{2}$ ansatz, variational principle yields a system
of implicit differential equations:

\begin{align}
 & \sum_{j}\left(\dot{\alpha}_{j,n}S_{ij}+\alpha_{j,n}S_{ij}K_{ij}\right)\nonumber \\
 & =-\text{i}\sum_{j,m}\alpha_{j,m}S_{ij}J_{nm}\nonumber \\
 & -\text{i}\sum_{j}\alpha_{j,n}S_{ij}\left(A_{ij}+B_{ij,n}+C_{ij,n}\right),
\end{align}
for each pair of indices $\left\{ i,n\right\} $, and

\begin{align}
 & \sum_{j,n}\left(\alpha_{i,n}^{\star}\dot{\alpha}_{j,n}S_{ij}\lambda_{j,kh}+P_{ij,n}\dot{\lambda}_{j,kh}+P_{ij,n}\lambda_{j,kh}K_{ij}\right)\nonumber \\
 & =-\text{i}\sum_{j,n,m}G_{ij,nm}\lambda_{j,kh}J_{nm}\nonumber \\
 & -\text{i}\sum_{j,n}P_{ij,n}\lambda_{j,kh}\left(A_{ij}+B_{ij,n}+C_{ij,n}\right)\nonumber \\
 & -\text{i}\sum_{j,n}P_{ij,n}\omega_{kh}^{g}\lambda_{j,kh}+\text{i}\sum_{j}P_{ij,k}f_{kh}\omega_{kh}^{e}\nonumber \\
 & -2\text{i}\sum_{j}P_{ij,k}\Delta\omega_{kh}\left(\lambda_{i,kh}^{\star}+\lambda_{j,kh}\right),
\end{align}
for pair of $\left\{ i,k,h\right\} $ indices, where we additionally
defined
\begin{align}
G_{ij,nm} & =\alpha_{i,n}^{\star}\alpha_{j,m}S_{ij},\\
P_{ij,n} & =G_{ij,nn},\\
A_{ij} & =\sum_{k,q}\omega_{kh}^{g}\lambda_{i,kh}^{\star}\lambda_{j,kh},\\
B_{ij,n} & =-\sum_{h}f_{nh}\omega_{nh}^{e}\left(\lambda_{i,nh}^{\star}+\lambda_{j,nh}\right),\\
C_{ij,n} & =\sum_{h}\Delta\omega_{nh}\left(1+\left(\lambda_{i,nh}^{\star}+\lambda_{j,nh}\right)^{2}\right).
\end{align}

For the $\text{D}_{2}$ ansatz, we once again can explicitly compute
equations of motions following TDVP, however, we do not have to, since
$\text{D}_{2}$ ansatz is a simplified version of $\text{mD}_{2}$
ansatz, when multiplicity number is set to $M=1$.

Calculation of linear response functions $S_{\text{abs/flor}}^{\left(1\right)}$
requires evaluation of two distinct coherent states. In the case of
$\text{D}_{2}$ and $\text{mD}_{2}$ ansatze, the overlap between
two distinct $a$ and $b$ coherent state are given b
\begin{equation}
\langle\lambda_{a}|\lambda_{b}\rangle=\exp\left(\lambda_{a}^{\ast}\lambda_{b}-\frac{1}{2}\left(|\lambda_{a}|^{2}+|\lambda_{b}|^{2}\right)\right).
\end{equation}
Meanwhile, overlap of two squeezed coherent states, as used in $\text{sqD}_{2}$
ansatz, is given by expression \citep{Chorosajev2017b}
\begin{align}
\langle\lambda_{a},\zeta_{a}|\lambda_{b},\zeta_{b}\rangle & =\frac{1}{\sqrt{\zeta_{ab}}}\exp\left(-\frac{\left|\lambda_{a}\right|^{2}+\left|\lambda_{b}\right|^{2}}{2}\right)\nonumber \\
 & \times\exp\left(\frac{\lambda_{a}^{*}\lambda_{b}}{\zeta_{ab}}+\frac{\lambda_{a}^{*2}}{2\zeta_{ab}}\eta_{ab}+\frac{\lambda_{b}^{2}}{2\zeta_{ab}}\eta_{ba}\right),
\end{align}
where
\begin{align}
\eta_{ab} & =e^{-\text{i}\theta_{b}}\cosh\left(r_{a}\right)\sinh\left(r_{b}\right)-e^{-\text{i}\theta_{a}}\cosh\left(r_{b}\right)\sinh\left(r_{a}\right),\\
\zeta_{ab} & =\cosh\left(r_{a}\right)\cosh\left(r_{b}\right)-e^{\text{i}\left(\theta_{b}-\theta_{a}\right)}\sinh\left(r_{a}\right)\sinh\left(r_{b}\right).
\end{align}

\bibliographystyle{rsc}
\bibliography{smD2}

\providecommand*{\mcitethebibliography}{\thebibliography}
\csname @ifundefined\endcsname{endmcitethebibliography}
{\let\endmcitethebibliography\endthebibliography}{}
\begin{mcitethebibliography}{43}
\providecommand*{\natexlab}[1]{#1}
\providecommand*{\mciteSetBstSublistMode}[1]{}
\providecommand*{\mciteSetBstMaxWidthForm}[2]{}
\providecommand*{\mciteBstWouldAddEndPuncttrue}
  {\def\EndOfBibitem{\unskip.}}
\providecommand*{\mciteBstWouldAddEndPunctfalse}
  {\let\EndOfBibitem\relax}
\providecommand*{\mciteSetBstMidEndSepPunct}[3]{}
\providecommand*{\mciteSetBstSublistLabelBeginEnd}[3]{}
\providecommand*{\EndOfBibitem}{}
\mciteSetBstSublistMode{f}
\mciteSetBstMaxWidthForm{subitem}
{(\emph{\alph{mcitesubitemcount}})}
\mciteSetBstSublistLabelBeginEnd{\mcitemaxwidthsubitemform\space}
{\relax}{\relax}

\bibitem[Valkunas \emph{et~al.}(2013)Valkunas, Abramavicius, and
  Man{\v{c}}al]{Valkunas2013a}
L.~Valkunas, D.~Abramavicius and T.~Man{\v{c}}al, \emph{{Molecular Excitation
  Dynamics and Relaxation}}, Wiley-VCH Verlag GmbH, 2013\relax
\mciteBstWouldAddEndPuncttrue
\mciteSetBstMidEndSepPunct{\mcitedefaultmidpunct}
{\mcitedefaultendpunct}{\mcitedefaultseppunct}\relax
\EndOfBibitem
\bibitem[Blankenship(2002)]{Blankenship2008}
R.~E. Blankenship, \emph{{Molecular Mechanisms of Photosynthesis}}, Blackwell
  Science Ltd, Oxford, UK, 2002\relax
\mciteBstWouldAddEndPuncttrue
\mciteSetBstMidEndSepPunct{\mcitedefaultmidpunct}
{\mcitedefaultendpunct}{\mcitedefaultseppunct}\relax
\EndOfBibitem
\bibitem[van Amerongen \emph{et~al.}(2000)van Amerongen, van Grondelle, and
  Valkunas]{Amerongen2010}
H.~van Amerongen, R.~van Grondelle and L.~Valkunas, \emph{{Photosynthetic
  Excitons}}, World Scientific, 2000\relax
\mciteBstWouldAddEndPuncttrue
\mciteSetBstMidEndSepPunct{\mcitedefaultmidpunct}
{\mcitedefaultendpunct}{\mcitedefaultseppunct}\relax
\EndOfBibitem
\bibitem[Davydov(1979)]{Davydov1979}
A.~S. Davydov, \emph{Physica Scripta}, 1979, \textbf{20}, 387--394\relax
\mciteBstWouldAddEndPuncttrue
\mciteSetBstMidEndSepPunct{\mcitedefaultmidpunct}
{\mcitedefaultendpunct}{\mcitedefaultseppunct}\relax
\EndOfBibitem
\bibitem[Scott(1991)]{Scott1991}
A.~C. Scott, \emph{Physica D: Nonlinear Phenomena}, 1991, \textbf{51},
  333--342\relax
\mciteBstWouldAddEndPuncttrue
\mciteSetBstMidEndSepPunct{\mcitedefaultmidpunct}
{\mcitedefaultendpunct}{\mcitedefaultseppunct}\relax
\EndOfBibitem
\bibitem[Zhao \emph{et~al.}(2022)Zhao, Sun, Chen, and Gelin]{Zhao2021}
Y.~Zhao, K.~Sun, L.~Chen and M.~Gelin, \emph{WIREs Computational Molecular
  Science}, 2022, \textbf{12}, e1589\relax
\mciteBstWouldAddEndPuncttrue
\mciteSetBstMidEndSepPunct{\mcitedefaultmidpunct}
{\mcitedefaultendpunct}{\mcitedefaultseppunct}\relax
\EndOfBibitem
\bibitem[Sun \emph{et~al.}(2010)Sun, Luo, and Zhao]{Sun2010b}
J.~Sun, B.~Luo and Y.~Zhao, \emph{Physical Review B - Condensed Matter and
  Materials Physics}, 2010, \textbf{82}, 014305\relax
\mciteBstWouldAddEndPuncttrue
\mciteSetBstMidEndSepPunct{\mcitedefaultmidpunct}
{\mcitedefaultendpunct}{\mcitedefaultseppunct}\relax
\EndOfBibitem
\bibitem[Choro{\v{s}}ajev \emph{et~al.}(2016)Choro{\v{s}}ajev, Rancova, and
  Abramavicius]{Chorosajev2016c}
V.~Choro{\v{s}}ajev, O.~Rancova and D.~Abramavicius, \emph{Physical Chemistry
  Chemical Physics}, 2016, \textbf{18}, 7966--7977\relax
\mciteBstWouldAddEndPuncttrue
\mciteSetBstMidEndSepPunct{\mcitedefaultmidpunct}
{\mcitedefaultendpunct}{\mcitedefaultseppunct}\relax
\EndOfBibitem
\bibitem[Jaku{\v{c}}ionis \emph{et~al.}(2018)Jaku{\v{c}}ionis,
  Choro{\v{s}}ajev, and Abramavi{\v{c}}ius]{Jakucionis2018b}
M.~Jaku{\v{c}}ionis, V.~Choro{\v{s}}ajev and D.~Abramavi{\v{c}}ius,
  \emph{Chemical Physics}, 2018, \textbf{515}, 193--202\relax
\mciteBstWouldAddEndPuncttrue
\mciteSetBstMidEndSepPunct{\mcitedefaultmidpunct}
{\mcitedefaultendpunct}{\mcitedefaultseppunct}\relax
\EndOfBibitem
\bibitem[Jaku{\v{c}}ionis \emph{et~al.}(2020)Jaku{\v{c}}ionis, Mancal, and
  Abramavi{\v{c}}ius]{Jakucionis2020a}
M.~Jaku{\v{c}}ionis, T.~Mancal and D.~Abramavi{\v{c}}ius, \emph{Physical
  Chemistry Chemical Physics}, 2020, \textbf{22}, 8952--8962\relax
\mciteBstWouldAddEndPuncttrue
\mciteSetBstMidEndSepPunct{\mcitedefaultmidpunct}
{\mcitedefaultendpunct}{\mcitedefaultseppunct}\relax
\EndOfBibitem
\bibitem[Sun \emph{et~al.}(2015)Sun, Gelin, Chernyak, and Zhao]{Sun2015a}
K.~W. Sun, M.~F. Gelin, V.~Y. Chernyak and Y.~Zhao, \emph{Journal of Chemical
  Physics}, 2015, \textbf{142}, 212448\relax
\mciteBstWouldAddEndPuncttrue
\mciteSetBstMidEndSepPunct{\mcitedefaultmidpunct}
{\mcitedefaultendpunct}{\mcitedefaultseppunct}\relax
\EndOfBibitem
\bibitem[Zhou \emph{et~al.}(2016)Zhou, Chen, Huang, Sun, Tanimura, and
  Zhao]{Zhou2016a}
N.~Zhou, L.~Chen, Z.~Huang, K.~Sun, Y.~Tanimura and Y.~Zhao, \emph{Journal of
  Physical Chemistry A}, 2016, \textbf{120}, 1562--1576\relax
\mciteBstWouldAddEndPuncttrue
\mciteSetBstMidEndSepPunct{\mcitedefaultmidpunct}
{\mcitedefaultendpunct}{\mcitedefaultseppunct}\relax
\EndOfBibitem
\bibitem[Choro{\v{s}}ajev \emph{et~al.}(2017)Choro{\v{s}}ajev,
  Mar{\v{c}}iulionis, and Abramavicius]{Chorosajev2017b}
V.~Choro{\v{s}}ajev, T.~Mar{\v{c}}iulionis and D.~Abramavicius, \emph{The
  Journal of Chemical Physics}, 2017, \textbf{147}, 074114\relax
\mciteBstWouldAddEndPuncttrue
\mciteSetBstMidEndSepPunct{\mcitedefaultmidpunct}
{\mcitedefaultendpunct}{\mcitedefaultseppunct}\relax
\EndOfBibitem
\bibitem[Jakucionis \emph{et~al.}(2022)Jakucionis, Gaiziunas, Sulskus, and
  Abramavicius]{Jakucionis2022}
M.~Jakucionis, I.~Gaiziunas, J.~Sulskus and D.~Abramavicius, \emph{The Journal
  of Physical Chemistry A}, 2022, \textbf{126}, 180--189\relax
\mciteBstWouldAddEndPuncttrue
\mciteSetBstMidEndSepPunct{\mcitedefaultmidpunct}
{\mcitedefaultendpunct}{\mcitedefaultseppunct}\relax
\EndOfBibitem
\bibitem[Zhou \emph{et~al.}(2016)Zhou, Chen, Huang, Sun, Tanimura, and
  Zhao]{Zhou2016}
N.~Zhou, L.~Chen, Z.~Huang, K.~Sun, Y.~Tanimura and Y.~Zhao, \emph{Journal of
  Physical Chemistry A}, 2016, \textbf{120}, 1562--1576\relax
\mciteBstWouldAddEndPuncttrue
\mciteSetBstMidEndSepPunct{\mcitedefaultmidpunct}
{\mcitedefaultendpunct}{\mcitedefaultseppunct}\relax
\EndOfBibitem
\bibitem[Wang \emph{et~al.}(2016)Wang, Chen, Zhou, and Zhao]{Wang2016}
L.~Wang, L.~Chen, N.~Zhou and Y.~Zhao, \emph{The Journal of Chemical Physics},
  2016, \textbf{144}, 024101\relax
\mciteBstWouldAddEndPuncttrue
\mciteSetBstMidEndSepPunct{\mcitedefaultmidpunct}
{\mcitedefaultendpunct}{\mcitedefaultseppunct}\relax
\EndOfBibitem
\bibitem[Chen \emph{et~al.}(2019)Chen, Gelin, and Domcke]{Chen2019a}
L.~Chen, M.~F. Gelin and W.~Domcke, \emph{Journal of Chemical Physics}, 2019,
  \textbf{150}, 24101\relax
\mciteBstWouldAddEndPuncttrue
\mciteSetBstMidEndSepPunct{\mcitedefaultmidpunct}
{\mcitedefaultendpunct}{\mcitedefaultseppunct}\relax
\EndOfBibitem
\bibitem[Jaku{\v{c}}ionis \emph{et~al.}(2022)Jaku{\v{c}}ionis, {\v{Z}}ukas, and
  Abramavi{\v{c}}ius]{MantasJakucionis2022}
M.~Jaku{\v{c}}ionis, A.~{\v{Z}}ukas and D.~Abramavi{\v{c}}ius, \emph{Physical
  Chemistry Chemical Physics}, 2022, \textbf{24}, 17665--17672\relax
\mciteBstWouldAddEndPuncttrue
\mciteSetBstMidEndSepPunct{\mcitedefaultmidpunct}
{\mcitedefaultendpunct}{\mcitedefaultseppunct}\relax
\EndOfBibitem
\bibitem[Abramavi{\v{c}}ius and Mar{\v{c}}iulionis(2018)]{Abramavicius2018e}
D.~Abramavi{\v{c}}ius and T.~Mar{\v{c}}iulionis, \emph{Lithuanian Journal of
  Physics}, 2018, \textbf{58}, 307--317\relax
\mciteBstWouldAddEndPuncttrue
\mciteSetBstMidEndSepPunct{\mcitedefaultmidpunct}
{\mcitedefaultendpunct}{\mcitedefaultseppunct}\relax
\EndOfBibitem
\bibitem[Bardeen(2014)]{Bardeen2014}
C.~J. Bardeen, \emph{Annual Review of Physical Chemistry}, 2014, \textbf{65},
  127--148\relax
\mciteBstWouldAddEndPuncttrue
\mciteSetBstMidEndSepPunct{\mcitedefaultmidpunct}
{\mcitedefaultendpunct}{\mcitedefaultseppunct}\relax
\EndOfBibitem
\bibitem[Schr{\"{o}}ter \emph{et~al.}(2015)Schr{\"{o}}ter, Ivanov, Schulze,
  Polyutov, Yan, Pullerits, and K{\"{u}}hn]{Schroter2015}
M.~Schr{\"{o}}ter, S.~Ivanov, J.~Schulze, S.~Polyutov, Y.~Yan, T.~Pullerits and
  O.~K{\"{u}}hn, \emph{Physics Reports}, 2015, \textbf{567}, 1--78\relax
\mciteBstWouldAddEndPuncttrue
\mciteSetBstMidEndSepPunct{\mcitedefaultmidpunct}
{\mcitedefaultendpunct}{\mcitedefaultseppunct}\relax
\EndOfBibitem
\bibitem[Steffen and Tanimura(2000)]{Steffen2013}
T.~Steffen and Y.~Tanimura, \emph{Journal of the Physical Society of Japan},
  2000, \textbf{69}, 3115--3132\relax
\mciteBstWouldAddEndPuncttrue
\mciteSetBstMidEndSepPunct{\mcitedefaultmidpunct}
{\mcitedefaultendpunct}{\mcitedefaultseppunct}\relax
\EndOfBibitem
\bibitem[Tanimura and Steffen(2000)]{Tanimura2013}
Y.~Tanimura and T.~Steffen, \emph{Journal of the Physical Society of Japan},
  2000, \textbf{69}, 4095--4106\relax
\mciteBstWouldAddEndPuncttrue
\mciteSetBstMidEndSepPunct{\mcitedefaultmidpunct}
{\mcitedefaultendpunct}{\mcitedefaultseppunct}\relax
\EndOfBibitem
\bibitem[Zhang \emph{et~al.}(2020)Zhang, Borrelli, and Tanimura]{Zhang2020}
J.~Zhang, R.~Borrelli and Y.~Tanimura, \emph{The Journal of Chemical Physics},
  2020, \textbf{152}, 214114\relax
\mciteBstWouldAddEndPuncttrue
\mciteSetBstMidEndSepPunct{\mcitedefaultmidpunct}
{\mcitedefaultendpunct}{\mcitedefaultseppunct}\relax
\EndOfBibitem
\bibitem[Hu \emph{et~al.}(1993)Hu, Paz, and Zhang]{Hu1993}
B.~L. Hu, J.~P. Paz and Y.~Zhang, \emph{Physical Review D}, 1993, \textbf{47},
  1576--1594\relax
\mciteBstWouldAddEndPuncttrue
\mciteSetBstMidEndSepPunct{\mcitedefaultmidpunct}
{\mcitedefaultendpunct}{\mcitedefaultseppunct}\relax
\EndOfBibitem
\bibitem[Xu \emph{et~al.}(2018)Xu, Liu, Zhang, and Yan]{Xu2018}
R.-X. Xu, Y.~Liu, H.-D. Zhang and Y.~Yan, \emph{The Journal of Chemical
  Physics}, 2018, \textbf{148}, 114103\relax
\mciteBstWouldAddEndPuncttrue
\mciteSetBstMidEndSepPunct{\mcitedefaultmidpunct}
{\mcitedefaultendpunct}{\mcitedefaultseppunct}\relax
\EndOfBibitem
\bibitem[Frenkel(1931)]{Frenkel1931}
J.~Frenkel, \emph{Physical Review}, 1931, \textbf{37}, 17--44\relax
\mciteBstWouldAddEndPuncttrue
\mciteSetBstMidEndSepPunct{\mcitedefaultmidpunct}
{\mcitedefaultendpunct}{\mcitedefaultseppunct}\relax
\EndOfBibitem
\bibitem[Werther and Gro{\ss}mann(2020)]{Werther2020}
M.~Werther and F.~Gro{\ss}mann, \emph{Physical Review B}, 2020, \textbf{101},
  174315\relax
\mciteBstWouldAddEndPuncttrue
\mciteSetBstMidEndSepPunct{\mcitedefaultmidpunct}
{\mcitedefaultendpunct}{\mcitedefaultseppunct}\relax
\EndOfBibitem
\bibitem[Glauber(1963)]{Glauber1963}
R.~J. Glauber, \emph{Physical Review}, 1963, \textbf{131}, 2766--2788\relax
\mciteBstWouldAddEndPuncttrue
\mciteSetBstMidEndSepPunct{\mcitedefaultmidpunct}
{\mcitedefaultendpunct}{\mcitedefaultseppunct}\relax
\EndOfBibitem
\bibitem[Wang \emph{et~al.}(2017)Wang, Fujihashi, Chen, and Zhao]{Wang2017b}
L.~Wang, Y.~Fujihashi, L.~Chen and Y.~Zhao, \emph{The Journal of Chemical
  Physics}, 2017, \textbf{146}, 124127\relax
\mciteBstWouldAddEndPuncttrue
\mciteSetBstMidEndSepPunct{\mcitedefaultmidpunct}
{\mcitedefaultendpunct}{\mcitedefaultseppunct}\relax
\EndOfBibitem
\bibitem[Xie \emph{et~al.}(2017)Xie, Zhong, Batchelor, and al]{Xie2017}
Q.~Xie, H.~Zhong, M.~T. Batchelor and al, \emph{Journal of Physics A:
  Mathematical and Theoretical}, 2017, \textbf{51}, 014001\relax
\mciteBstWouldAddEndPuncttrue
\mciteSetBstMidEndSepPunct{\mcitedefaultmidpunct}
{\mcitedefaultendpunct}{\mcitedefaultseppunct}\relax
\EndOfBibitem
\bibitem[Mukamel(1995)]{Mukamel1995a}
S.~Mukamel, \emph{{Principles of nonlinear optical spectroscopy}}, Oxford
  University Press, 1995\relax
\mciteBstWouldAddEndPuncttrue
\mciteSetBstMidEndSepPunct{\mcitedefaultmidpunct}
{\mcitedefaultendpunct}{\mcitedefaultseppunct}\relax
\EndOfBibitem
\bibitem[Balevi{\v{c}}ius \emph{et~al.}(2015)Balevi{\v{c}}ius, Valkunas, and
  Abramavicius]{Balevicius2015c}
V.~Balevi{\v{c}}ius, L.~Valkunas and D.~Abramavicius, \emph{The Journal of
  Chemical Physics}, 2015, \textbf{143}, 074101\relax
\mciteBstWouldAddEndPuncttrue
\mciteSetBstMidEndSepPunct{\mcitedefaultmidpunct}
{\mcitedefaultendpunct}{\mcitedefaultseppunct}\relax
\EndOfBibitem
\bibitem[Zhan \emph{et~al.}(2009)Zhan, Zhang, Li, and Chung]{Zhan2009}
Z.~H. Zhan, J.~Zhang, Y.~Li and H.~S. Chung, \emph{IEEE transactions on
  systems, man, and cybernetics. Part B, Cybernetics : a publication of the
  IEEE Systems, Man, and Cybernetics Society}, 2009, \textbf{39},
  1362--1381\relax
\mciteBstWouldAddEndPuncttrue
\mciteSetBstMidEndSepPunct{\mcitedefaultmidpunct}
{\mcitedefaultendpunct}{\mcitedefaultseppunct}\relax
\EndOfBibitem
\bibitem[{K Mogensen} and {N Riseth}(2018)]{mogensen2018optim}
P.~{K Mogensen} and A.~{N Riseth}, \emph{Journal of Open Source Software},
  2018, \textbf{3}, 615\relax
\mciteBstWouldAddEndPuncttrue
\mciteSetBstMidEndSepPunct{\mcitedefaultmidpunct}
{\mcitedefaultendpunct}{\mcitedefaultseppunct}\relax
\EndOfBibitem
\bibitem[Lim \emph{et~al.}(2015)Lim, Pale{\v{c}}ek, Caycedo-Soler, Lincoln,
  Prior, von Berlepsch, Huelga, Plenio, Zigmantas, and Hauer]{Lim2015a}
J.~Lim, D.~Pale{\v{c}}ek, F.~Caycedo-Soler, C.~N. Lincoln, J.~Prior, H.~von
  Berlepsch, S.~F. Huelga, M.~B. Plenio, D.~Zigmantas and J.~Hauer,
  \emph{Nature Communications}, 2015, \textbf{6}, 7755\relax
\mciteBstWouldAddEndPuncttrue
\mciteSetBstMidEndSepPunct{\mcitedefaultmidpunct}
{\mcitedefaultendpunct}{\mcitedefaultseppunct}\relax
\EndOfBibitem
\bibitem[Christensson \emph{et~al.}(2011)Christensson, Milota, Hauer, Sperling,
  Bixner, Nemeth, and Kauffmann]{Christensson2011}
N.~Christensson, F.~Milota, J.~Hauer, J.~Sperling, O.~Bixner, A.~Nemeth and
  H.~F. Kauffmann, \emph{Journal of Physical Chemistry B}, 2011, \textbf{115},
  5383--5391\relax
\mciteBstWouldAddEndPuncttrue
\mciteSetBstMidEndSepPunct{\mcitedefaultmidpunct}
{\mcitedefaultendpunct}{\mcitedefaultseppunct}\relax
\EndOfBibitem
\bibitem[Bondarenko \emph{et~al.}(2020)Bondarenko, Jansen, and
  Knoester]{Bondarenko2020}
A.~S. Bondarenko, T.~L.~C. Jansen and J.~Knoester, \emph{The Journal of
  Chemical Physics}, 2020, \textbf{152}, 194302\relax
\mciteBstWouldAddEndPuncttrue
\mciteSetBstMidEndSepPunct{\mcitedefaultmidpunct}
{\mcitedefaultendpunct}{\mcitedefaultseppunct}\relax
\EndOfBibitem
\bibitem[Hestand and Spano(2018)]{Hestand2018}
N.~J. Hestand and F.~C. Spano, \emph{Chemical Reviews}, 2018, \textbf{118},
  7069--7163\relax
\mciteBstWouldAddEndPuncttrue
\mciteSetBstMidEndSepPunct{\mcitedefaultmidpunct}
{\mcitedefaultendpunct}{\mcitedefaultseppunct}\relax
\EndOfBibitem
\bibitem[Zeng and Yao(2022)]{Zeng2022a}
J.~Zeng and Y.~Yao, \emph{Journal of Chemical Theory and Computation}, 2022,
  \textbf{18}, 1255--1263\relax
\mciteBstWouldAddEndPuncttrue
\mciteSetBstMidEndSepPunct{\mcitedefaultmidpunct}
{\mcitedefaultendpunct}{\mcitedefaultseppunct}\relax
\EndOfBibitem
\bibitem[Beck(2000)]{Beck2000a}
M.~Beck, \emph{Physics Reports}, 2000, \textbf{324}, 1--105\relax
\mciteBstWouldAddEndPuncttrue
\mciteSetBstMidEndSepPunct{\mcitedefaultmidpunct}
{\mcitedefaultendpunct}{\mcitedefaultseppunct}\relax
\EndOfBibitem
\bibitem[Worth and Burghardt(2003)]{Worth2003a}
G.~A. Worth and I.~Burghardt, \emph{Chemical Physics Letters}, 2003,
  \textbf{368}, 502--508\relax
\mciteBstWouldAddEndPuncttrue
\mciteSetBstMidEndSepPunct{\mcitedefaultmidpunct}
{\mcitedefaultendpunct}{\mcitedefaultseppunct}\relax
\EndOfBibitem
\bibitem[Worth \emph{et~al.}(2008)Worth, Meyer, K{\"{o}}ppel, Cederbaum, and
  Burghardt]{Worth2009}
G.~A. Worth, H.-D. Meyer, H.~K{\"{o}}ppel, L.~S. Cederbaum and I.~Burghardt,
  \emph{International Reviews in Physical Chemistry}, 2008, \textbf{27},
  569--606\relax
\mciteBstWouldAddEndPuncttrue
\mciteSetBstMidEndSepPunct{\mcitedefaultmidpunct}
{\mcitedefaultendpunct}{\mcitedefaultseppunct}\relax
\EndOfBibitem
\end{mcitethebibliography}

\end{document}